\documentclass[12pt]{iopart}

\usepackage{graphicx}
\usepackage{natbib}
\usepackage{ulem}
\usepackage{amssymb}

\newcommand{\x}[1]{^{(#1)}}                                     %
\newcommand{\R}{\bar{\omega}_{r0}}                              %
\newcommand{\T}{\bar{\omega}_{\theta0}}                         %
\newcommand{\be}{\begin{equation}}                              %
\newcommand{\ee}{\end{equation}}                                %
\newcommand{\bea}{\begin{eqnarray}}                             %
\newcommand{\eea}{\end{eqnarray}}                               %
\newcommand{\nn}{\nonumber}
\newcommand{\bx}{\bar{x}} 
\newcommand{\by}{\bar{y}} 
\begin{document}

\title[Epicyclic oscillations of non-slender tori]{Epicyclic oscillations of non-slender fluid tori around Kerr black holes}

\author{Odele Straub$^1$ and Eva {\v S}r{\'a}mkov{\'a}$^2$}

\address{$^1$Copernicus Astronomical Centre PAN, Bartycka 18, 00-716 Warsaw, Poland}
\address{$^2$Department of Physics, Silesian University in Opava, Bezru{\v c}ovo n{\'a}m. 13, 746-01 Opava, Czech Republic}

\eads{\mailto{odele@camk.edu.pl}, \mailto{sram\_eva@centrum.cz}}


\begin{abstract}
Considering epicyclic oscillations of pressure-supported perfect fluid tori orbiting Kerr black holes we examine non-geodesic (pressure) effects on the epicyclic modes properties. Using a perturbation method we derive fully general relativistic formulas for eigenfunctions and eigenfrequencies of the radial and vertical epicyclic modes of a slightly non-slender, constant specific angular momentum torus up to second-order accuracy with respect to the torus thickness. The behaviour of the axisymmetric and lowest-order ($m=\pm 1$) non-axisymmetric epicyclic modes is investigated. For an arbitrary black hole spin we find that, in comparison with the (axisymmetric) epicyclic frequencies of free test particles, non-slender tori receive negative pressure corrections and exhibit thus lower frequencies. Our findings are in qualitative agreement with the results of a recent pseudo-Newtonian study of analogous problem defined within the Paczy{\'n}ski-Wiita potential. Implications of our results on the high-frequency QPO models dealing with epicyclic oscillations are addressed.
\end{abstract}

\vspace{2pc}
\submitto{\CQG}

\section{Introduction}
\label{sec:intro}
Oscillations of black hole accretion discs have been studied extensively in various astrophysical contexts. Investigations of hydrodynamic oscillation modes of geometrically thin accretion discs revealed three fundamental, discoseismic classes of modes: acoustic pressure p-modes, gravity g-modes and corrugation c-modes. \citet{kat98, wag99, kat01a} and \citet{wag01} give comprehensive reviews on the subject of relativistic 'discoseismology'. Geometrically thick discs (tori) have been examined rather less and mainly in matters of their stability \citep[e.g.][]{pap84, koj86}. Only recently oscillatory modes of fluid tori were explored in more detail in several numerical studies \citep [e.g.][] {zan03, rez03a, rub05a, rub05b, sra07, mon07} and in a purely analytic work by \citet{bla06}. The latter presents a thorough analysis of oscillatory modes of relativistic slender tori.

The analysis of disc oscillation modes is motivated by observations of quasi-periodic oscillations (QPOs) in the light curves of Galactic low-mass X-ray binaries \citep[see][for reviews]{mcc03, van04}. In particular the understanding of high frequency (HF) QPOs, which are presumably a strong gravity phenomenon, would provide a deeper insight into the innermost regions of accretion discs and the very nature of compact objects. Black hole HF QPOs occur at frequencies that are constant in time and characteristic for a particular source. If more than one HF QPO is detected in a given system, the frequencies typically appear in ratios of small natural numbers, whereas in most cases the ratio is close to 3:2 \citep{abr01, mcc03}. 

Several models explain HF QPOs in terms of disc (or blob) oscillations. \citet{ste98} for instance consider an orbiting hot spot and propose that HF QPOs arise due to a modulation of the spot radiation by its precessional motion. Pointing out the observed rational ratios of HF QPO pairs \citet{klu00} and \citet{abr01} suggest an underlying non-linear resonance between some modes of accretion disc oscillation. In the resonance model the mode pair is commonly represented by the radial and vertical epicyclic oscillation. The importance of epicyclic oscillations is also stressed by \citet{kat01b} who attributes the origin of HF QPOs to non-axisymmetric g-mode oscillations. The corotation resonance in a disc which is deformed by a warp would be responsible for the excitation of g-modes and general relativistic effects trap them near the inner edge of the accretion disc \citep{kat03, kat04}. HF QPOs could also result from acoustic p-mode oscillations of a small accretion torus orbiting close to the black hole \citep{rez03b}. Recently \citet{bla06} discussed the possibility that the vertical epicyclic and the lowest-order (acoustic) breathing mode of a relativistic slender torus might represent the two black hole HF QPOs in 3:2 ratio.

The main interest of our work lies in the epicyclic modes of oscillations. Most models that are dealing with them consider geodesic flows and are based on free test particles. However, non-geodesic effects related to e.g. magnetic fields, viscosity or pressure forces may play a certain role in this concern. The aim of this paper is to investigate such non-geodesic effects on the two epicyclic modes by means of a pressure-supported perfect fluid torus and to find the consequential pressure corrections to the mode eigenfunctions and eigenfrequencies.

For an infinitely slender torus, the frequencies of epicyclic oscillations are consistent with the epicyclic frequencies of free test particles on a circular orbit in the equatorial plane, where the torus pressure exhibits a maximum value \citep{abr06}. It has been shown that the epicyclic modes may be retained also for thicker tori \citep{bla07}. Numerical simulations \citep{rez03a, rub05a, rub05b, sra07} as well as analytic calculations in pseudo-Newtonian approximation \citep{bla07} showed that with growing torus thickness the (axisymmetric) epicyclic frequencies decrease. What we have in mind here is a slightly non-slender torus (with a radial extent that is very small in comparison to its distance from the central object) in hydrostatic equilibrium in Kerr spacetime. We perform a second-order perturbation analysis of the eigenfunctions and eigenfrequencies of both modes with respect to the torus thickness and derive exact analytic formulas for the pressure corrections. 

\citet{bla07} have studied an analogous problem in the framework of Newtonian physics using the pseudo-Newtonian potential of \citet{pac80} to model the relevant general relativistic effects. Our work represents a generalisation of their results into Kerr geometry. Like \citet{bla07} we assume a non-self-gravitating, non-magnetic, stationary torus with a constant specific angular momentum distribution. We neglect self-gravity because the mass of a black hole exceeds the mass of a torus many times over. Although effects of magnetic fields are neither negligible nor irrelevant to our problem, we study here a purely hydrodynamical case. Our calculations are only valid as long as magnetic fields are unimportant, i.e., as long as the torus pressure dominates the magnetic pressure. The stability of these hydrodynamic modes in presence of magnetohydrodynamical turbulence is an issue that needs yet to be investigated. The above assumptions are supported by numerical simulations. \citet{pro03a} show for instance in an inviscid hydrodynamical simulation that the inner axisymmetric accretion flow settles into a pressure-rotation supported torus with constant specific angular momentum. Once magnetic fields are introduced the angular momentum distribution gets a different profile, torus-like configurations are, however, still seen as ''inner tori'' in global MHD simulations \citep[e.g.][]{mac06, fra08}. Although the described torus setup is not the most likely to be found in nature, we think it is reasonable to assume such a configuration as a first approximation.

In this work we focus on a mathematical description of the problem, whereas the astrophysical applications shall be presented separately. The paper is outlined in the following manner: In sections \ref{sec:equilibrium} and \ref{sec:PP} we give a brief introduction to the problem writing down the equations that describe the relativistic equilibrium tori and the relativistic Papaloizou-Pringle equation. Then, in section \ref{sec:calculation}, we describe the perturbation method and derive formulas for the radial and vertical epicyclic mode eigenfunctions and eigenfrequencies. The results are presented for different values of the black hole spin parameter in section \ref{sec:figures} and discussed in section \ref{sec:conclusions}.

\section{Equilibrium configuration}
\label{sec:equilibrium}

Consider an axisymmetric, non-self-gravitating perfect fluid torus in hydrostatic equilibrium on the background of the Kerr geometry. The flow of fluid is stationary and in a state of pure rotation. Generally, the line element of a stationary, axially symmetric spacetime is given in Boyer-Lindquist coordinates ($t,r,\theta,\phi$) by
\be
ds^2 = g_{tt}dt^2 + 2g_{t\phi}dt d\phi + g_{rr}dr^2 + g_{\theta\theta}d\theta^2 + g_{\phi\phi}d\phi^2.
\ee
We take the ($- + + +$) signature and units where $c=G=M=1$. The explicit expressions for the covariant and contravariant coefficients of the Kerr metric then write 
\bea
&g_{tt} = -\left(1-\frac{2r}{\Sigma}\right),     \qquad &g^{tt} = -\frac{\Xi} {\Sigma \Delta}, \nn\\
&g_{t\phi} = -\frac{2ar}{\Sigma}\sin^2\theta,  \qquad &g^{t\phi} = -\frac{2ar}{ \Sigma\Delta}, \nn\\
&g_{rr} = \frac{\Sigma}{\Delta},                 \qquad &g^{rr} = \frac{\Delta}{\Sigma}, \nn\\
&g_{\theta\theta} = \Sigma,                      \qquad &g^{\theta\theta} = \frac{1}{\Sigma}, \nn\\
&g_{\phi\phi} = \left(r^2+a^2+\frac{2a^2r}{\Sigma}\sin^2\theta\right)\sin^2\theta,  \qquad &g^{\phi\phi} = \frac{\Delta-a^2\sin^2\theta}{\Sigma\Delta\sin^2\theta},
\label{eq:metric}
\eea
where $\Sigma\equiv r^2+a^2\cos^2\theta$, $\Delta\equiv r^2-2r+a^2$, $\Xi\equiv(r^2+a^2)^2-a^2\Delta\sin^2\theta$ and $M$ is the mass and $a$ the specific angular momentum (spin) of the black hole.

Because the flow is assumed to be purely azimuthal, the four-velocity  has only two non-zero components, $u^{\mu}=(u^t,0,0,u^{\phi})$. One may derive the fluid specific angular momentum $l$, angular velocity $\Omega$,  specific energy $\mathcal{E}$ and the contravariant $t$-component of the four-velocity, often denoted as $A$, in the form
\bea
l \equiv    - & \frac{u_\phi}{u_t}   & = -\frac{g_{t\phi}+\Omega g_{\phi\phi}}{g_{tt}+\Omega g_{t\phi}},\\
\label{eq:l}
\Omega \equiv & \frac{u^\phi}{u^t}   & =\frac{g^{t\phi}-lg^{\phi\phi}}{g^{tt}-lg^{t\phi}},\\
\label{eq:Omega}
\mathcal{E} \equiv - & u_t           & = (-g^{tt}+2lg^{t\phi}-l^2g^{\phi\phi})^{-1/2},\\
\label{eq:E}
A \equiv             & u^t           & = (-g_{tt}-2\Omega g_{t\phi}-\Omega^2g_{\phi\phi})^{-1/2}.
\label{eq:A}
\eea

The perfect fluid is characterised by the stress-energy tensor $T^{\mu\nu}=(p+e)u^{\mu}u^{\nu} + p g^{\mu\nu}$. We restrict our consideration to polytropic flows such that, measured in the fluid's rest frame, pressure $p$, internal energy density $e$ and rest mass density $\rho$ are related by $p = K\rho^{(n+1)/n}$ and $e = np + \rho$, where $n$ is the polytropic index and $K$ the polytropic constant. 

We assume the specific angular momentum to be constant throughout the torus, i.e., $l(r,\theta) \equiv l_0 = const$. Such a configuration is governed by the relativistic Euler equation which may be written as
\be
-\frac{ \partial_{\mu} p}{p+e} = \partial_{\mu} \left(\ln\mathcal{E}\right),  \qquad   \mu \in \{r,\theta\}.
\label{eq:euler}
\ee
Introducing the enthalpy  $H\equiv \int{dp/(p+e)}$, the integration of \eref{eq:euler} leads for a barotropic fluid, for which $p=p(e)$, to the following form of the Bernoulli equation 
\be
H + \ln\mathcal{E} = const.
\label{eq:bernoulli}
\ee
that determines the structure of the torus in the $r-\theta$ plane. The subscript zero refers to the special location $r=r_0$ in the equatorial plane where the pressure gradients vanish ($p$ has a maximal value) and the fluid moves along a geodesic line. 

For a small torus cross-section, when the adiabatic sound speed defined at the pressure maximum $c^2_{s0} = (n+1)p_0/(n\rho_0)$ satisfies $c^2_{s0} << c^2$, one may write $H \approx (n+1) p/\rho$ \citep{abr06}. Following \citet{abr06} and \citet{bla07} we introduce the function $f(r,\theta)$, which takes constant values at the isobaric and isodensity surfaces, by
\be
\frac{p}{\rho}=\frac{p_0}{\rho_0}f(r,\theta).
\label{eq:f def}
\ee
Form the Bernoulli equation (\ref{eq:bernoulli}) with the constant evaluated at the pressure maximum $r_0$ and
the above form of $H$  we get
\be
f = 1-\frac{1}{n c_{s0}^2}\left(\ln\mathcal{E}-\ln\mathcal{E}_0\right).
\label{eq:f}
\ee

\subsection{Epicyclic oscillations}

It is advantageous to introduce the effective potential
\be
U = g^{tt}-2l_{0}g^{t\phi}+l_{0}^2g^{\phi\phi}
\label{eq:U}
\ee
that has its minimum at the torus pressure maximum $r_0$. A small perturbation of a test particle orbiting on a geodesic line $r=r_0$ with $l=l_0$ results in radial and vertical epicyclic oscillations around the equilibrium point $r_0$ at a radial $\omega_{r0}$ and vertical $\omega_{\theta0}$ epicyclic frequency given by (e.g., \cite{abr06})
\be
\omega_{r0}^2 = \frac{1}{2}\left(\frac{\mathcal{E}^2}{A^2g_{rr}}\frac{\partial^2 U}{\partial r^2}\right)_0 
\qquad  \textnormal{and} \qquad
\omega_{\theta0}^2= \frac{1}{2}\left(\frac{\mathcal{E}^2}{A^2g_{\theta\theta}}\frac{\partial^2 U}{\partial \theta^2}\right)_0.
\label{eq:testp freq}
\ee
In Kerr geometry \eref{eq:metric} the above definitions lead to \citep{ali81, now98, tor05b}
\be
\fl\omega_{r0}^2 = \Omega_0^2 \left(1-\frac{6}{r_0}+\frac{8a}{r_0^{3/2}}-\frac{3a^2}{r_0^2}\right)
\qquad \textnormal{and} \qquad
\omega_{\theta0}^2 = \Omega_0^2 \left(1-\frac{4a}{r_0^{3/2}}+\frac{3a^2}{r_0^2}\right),
\label{eq:RT}
\ee
where $\Omega_0$ is the angular velocity at the pressure maximum $r_0$ that in Kerr geometry reads $\Omega_0=1/(r_0^{3/2}+a)$.

In order to investigate the behaviour of the equipotential function $f$ in close vicinity of the equilibrium point $r_0$, \citet{abr06} introduced local coordinates,
\be
x = \sqrt{g_{rr0}} \left(\frac{r - r_0}{r_0}\right) \qquad  \textnormal{and} \qquad 
y = \sqrt{g_{\theta\theta0}}\left(\frac{\pi/2 - \theta}{r_0}\right),
\label{eq:xy}
\ee
satisfying $x = y = 0$ at $r_0$.
For small $x$ and $y$, and a constant specific angular momentum torus, the equipotential function $f$ can be expressed as
\be
f = 1-\frac{1}{\beta^2}\left(\R^2 x^2 + \T^2 y^2\right),
\label{eq:f xy}
\ee
where $\R \equiv \omega_{r0}/\Omega_0$, $\T \equiv \omega_{\theta0}/\Omega_0$, and $\beta$ is a dimensionless parameter given by
\be
\beta^2 \equiv \frac{2nc_{s0}^2}{r_0^2 A_0^2 \Omega_0^2}
\label{eq:beta}
\ee
which determines the thickness of the torus. 

\Eref{eq:f xy} describes the equipotential function in the vicinity of $r_0$ in terms of the test particle epicyclic frequencies and is congruent with the formula derived in Newtonian theory (see equation (9) in \citet{bla07}). In the slender torus limit $\beta \rightarrow 0$, the torus reduces to an infinitesimally slender ring at the pressure maximum $r_0$. In a Newtonian $1/r$ potential, $\omega_{r0} = \omega_{\theta0} = \Omega_0$ and the slender torus cross-section has a circular shape. In the general case, however, $\omega_{r0} \neq \omega_{\theta0}$ and the isobaric surfaces are ellipses with semi-axes being in the ratio of the two epicyclic frequencies.

\section{Perturbation equation}
\label{sec:PP}

We consider small linear perturbations around the axisymmetric and stationary torus equilibrium with azimuthal and time dependence in the form $\propto\exp[\rmi(m\phi-\omega t)]$. The differential equation governing such perturbations for constant specific angular momentum tori was in Newtonian theory derived  by \citet{pap84} where it was expressed in terms of a scalar variable $W$. \citet{abr06} recently derived the general relativistic form of the Papaloizou-Pringle equation in terms of 
\be
W = -\frac{\delta p}{A\rho(\omega - m\Omega)},
\ee
which is related to the Eulerian perturbation in the four-velocity as 
\be
\delta u_{\mu} = \frac{\rmi\rho}{p+e} \partial_{\mu}W,  \qquad   \mu \in \{r,\theta\}.
\label{eq:velocity}
\ee
The relativistic Papaloizou-Pringle equation writes
\bea
\frac{1}{(-g)^{1/2}}\left\{\partial_{\mu}\left[(-g)^{1/2}g^{\mu\nu}f^n\partial_{\nu}W\right]\right\}&-&\left(m^2g^{\phi\phi} - 2m\omega g^{t\phi} + \omega^2 g^{tt}\right)f^n W\nn\\
&& = -\frac{2n\mathcal{A}(\bar{\omega}-m\bar{\Omega})^2}{\beta^2 r^2_0}f^{n-1}W
\label{eq:PP}
\eea
where $\{\mu,\nu\}\in\{r,\theta\}$, $\mathcal{A}\equiv A^2/A^2_0$, $\bar{\Omega}\equiv\Omega/\Omega_{0}$, $\bar{\omega} \equiv \omega/\Omega_{0}$ and $g$ denotes the determinant of the metric.

Following \citet{bla07} we write (\ref{eq:PP}) as
\be
\hat{L}W = -2n\mathcal{A}\left(\bar{\omega}-m\bar{\Omega}\right)^2 W,
\label{eq:eigenvalue}
\ee
where $\hat{L}$ is a linear operator given by
\bea
\hat{L}&=&[\frac{1}{\sqrt{-g}}\partial_{\mu}(\sqrt{-g})g^{\mu\nu}f\partial_{\nu}+\partial_{\mu}(g^{\mu\nu})f\partial_{\nu}+g^{\mu\nu}n\partial_{\mu} f\partial_{\nu}+g^{\mu\nu}f\partial^2_{\nu}\nn\\
&&-\left(m^2g^{\phi\phi}-2m\omega g^{t\phi}+\omega^{2}g^{tt}\right)f]\beta^2r^2_0.
\label{eq:L}
\eea

\section{Expanding the relativistic Papaloizou-Pringle equation about the slender torus limit}
\label{sec:calculation}

We now use a perturbation method to derive the expressions for eigenfunctions and eigenfrequencies of the epicyclic modes for thicker tori ($\beta>0$). 

We start by transforming all variables to local coordinates, $\bx=x/\beta$ and $\by=y/\beta$, measured from the equilibrium point. Then we expand $\bar{\omega}$, $W$, $\mathcal{A}$, $\bar{\Omega}$, $f$ and  $\hat{L}$ into a power series in $\beta$ by writing
\be
Q =Q\x{0}+Q\x{1}\beta+Q\x{2}\beta^2+\ldots\\ \,\, \textnormal{where} \,\,\,\,  Q \in \{\bar{\omega}, W, \mathcal{A}, \bar{\Omega}, f, \hat{L}\}.
\label{eq:Q}
\ee
Substituting all variables \eref{eq:Q} into the perturbation equation \eref{eq:eigenvalue} and comparing terms of the same order in $\beta$ we obtain formulas for the respective corrections to eigenfunctions and eigenfrequencies of the desired modes.

\subsection{Slender torus limit}

In the slender torus limit ($\beta \rightarrow 0$) the relativistic Papaloizou-Pringle equation (\ref{eq:PP}) reduces to
\bea
f\x{0} \left(\frac{\partial^2W\x{0}}{\partial \bx^2} + \frac{\partial^2W\x{0}}{\partial \by^2} \right) + n \left(\frac{\partial f\x{0}}{\partial \bx} \frac{\partial W\x{0}}{\partial \bx} + \frac{\partial f\x{0}}{\partial \by} \frac{\partial W\x{0}}{\partial \by} \right) = \nn \\
 -2n \mathcal{A}\x{0}(\bar{\omega}\x{0}-m\bar{\Omega}\x{0})^2 W\x{0}.
\label{eq:PPslender}
\eea
It represents the zeroth-order of \eref{eq:eigenvalue} and may be written in operator form,
\be
\hat{L}\x{0}W\x{0}=-2n\mathcal{A}\x{0}\sigma^2 W\x{0},
\label{eq:PP0}
\ee
where $\sigma \equiv \bar{\omega}\x{0} - m\bar{\Omega}\x{0}$ denotes the zeroth-order eigenfrequency in the corotating frame scaled with the orbital velocity $\Omega_0$. The encountered zeroth-order expansion terms are
\bea
\mathcal{A}\x{0} &=& 1,\\
\bar{\Omega}\x{0} &=& 1,\\
f\x{0} &=&  1- \R^2 \bx^2 - \T^2 \by^2,\\
\hat{L}\x{0} &=& f\x{0}\frac{\partial^2}{\partial \bx^2} + f\x{0}\frac{\partial^2}{\partial \by^2} + n\frac{\partial f\x{0}}{\partial \bx}\frac{\partial}{\partial \bx} + n\frac{\partial f\x{0}}{\partial \by}\frac{\partial}{\partial \by}.
\eea

Equations \eref{eq:PPslender} and \eref{eq:PP0} are of the same form as the Newtonian slender torus limit of the Papaloizou-Pringle equation \citep{abr06, bla07}. They represent an eigenvalue equation for $\sigma$ with $\hat{L}\x{0}$ being a self-adjoint operator with respect to the inner product
\be
\left\langle W_a\x{0}|W_b\x{0}\right\rangle = \int\int\left(f\x{0}\right)^{n-1} W_a^{(0)\ast} W_b\x{0}d\bx d\by = \delta_{ab},
\label{eq:inner product}
\ee
where the integrals are taken over the slender torus cross-section where $f\x{0} \geq 0$. This implies that the eigenvalues $\sigma$ are real numbers and the zeroth-order eigenfunctions $W\x{0}$ form a complete orthonormal set. Therefore, any function defined over the torus cross-section may be expanded in terms of $W\x{0}$. 

The explicit expressions for eigenfunctions and eigenfrequencies of the complete set of normal modes of a constant specific angular momentum slender torus in Newtonian potential, where $\R = \T = 1$, were given in \citet{bla85}. Recently, \citet{bla06} derived the eigenfunctions and eigenfrequencies of the lowest-order modes of a general relativistic slender torus, where $\R \neq \T$, for arbitrary specific angular momentum distribution. The eigenfunctions and eigenfrequencies of the simplest modes of the relativistic torus with constant specific angular momentum which are relevant to our calculations are specified in table \ref{tab:eigenfunctions}.

In the slender torus limit the two epicyclic modes correspond to $i=1$ and $i=2$ from table \ref{tab:eigenfunctions}. The corresponding eigenfunctions take the form
\be
W_r\x{0} \equiv W_1\x{0} = a_1 \bx e^{\rmi(m\phi-\omega_1\x{0}t)} \quad \textnormal{and} \quad
W_\theta\x{0} \equiv W_2\x{0} = a_2 \by e^{\rmi(m\phi-\omega_2\x{0}t)}.
\label{eq:W1W2}
\ee
They describe the global, incompressible modes in which the entire torus moves in purely radial ($W_r\x{0}$) or purely vertical ($W_\theta\x{0}$) direction at frequencies which are, in the corotating frame, consistent with the epicyclic frequencies of free test particles, $\omega_{r0} = \sigma_1 \Omega_0 = \omega_1\x{0}-m \Omega \x{0}$ and $\omega_{\theta0}=\sigma_2 \Omega_0=\omega_2\x{0}-m \Omega \x{0}$.

\subsection{Non-slender torus}

\subsubsection{First-order corrections}
\label{sec:first}

Expanding the Papaloizou-Pringle equation \eref{eq:eigenvalue} to first-order in $\beta$ we arrive at \footnote{We use the index $i$ to label the modes of interest. It takes values $i=1$ for the radial or $i=2$ for the vertical epicyclic mode.}
\bea
\hat{L}\x{0}W_i\x{1}+\hat{L}\x{1}W_i\x{0} &=& 2n(2m\sigma_i\bar{\Omega}\x{1} - \sigma_i^2\mathcal{A}\x{1} - 2\sigma_i\bar{\omega}_i\x{1})W_i\x{0}\nn\\
&& -2n\sigma_i^2W_i\x{1}.
\label{eq:PP1}
\eea
The perturbed basis of eigenfunctions $W_i\x{1}$ may now be expressed in terms of the orthonormal zeroth-order basis as
\be
W_i\x{1}=\sum_{j}{b_{ij}W_j\x{0}}.
\label{eq:W1}
\ee
The $b_{ij}$ coefficients may be determined by taking the inner product of (\ref{eq:PP1}) with a zeroth-order eigenfunction $W_{k}\x{0}$. If the subscript $k$ refers to a different mode from the one we are interested in (i.e., $k\neq i$) we find
\be
b_{ij}=\frac{\left\langle W_j\x{0}|\hat{L}\x{1}-4nm\sigma_i\bar{\Omega}\x{1}+2n\sigma_i^2\mathcal{A}\x{1}|W_i\x{0}\right\rangle}{2n(\sigma_j^2-\sigma_i^2)}.
\label{eq:bij}
\ee
Whereas if $k$ refers to either of the two epicyclic modes ($k=i$) we obtain the formula for the first-order correction to the radial ($i=1$) or vertical ($i=2$) eigenfrequency,
\be
\bar{\omega}_i\x{1}=-\frac{\left\langle W_i\x{0}|\hat{L}\x{1}-4nm\sigma_i\bar{\Omega}\x{1} +2n\sigma_i^2\mathcal{A}\x{1}|W_i\x{0}\right\rangle}{4n\sigma_i}.
\label{eq:omega1}
\ee

The first-order expansion terms of $\mathcal{A}$, $\bar{\Omega}$, $f$ and $\hat{L}$ have a ($\bx,\by$) dependence in the form
\bea
\mathcal{A}\x{1} &&= \mathcal{A}_{11} \bx, \label{eq:A1} \\
\bar{\Omega}\x{1} &&= \Omega_{11} \bx,\\
\label{eq:Omega1}
f\x{1} &&= f_{11} \bx^3 + f_{12} \bx \by^2,\\
\label{eq:f1}
\hat{L}\x{1} &&= \left(L_{101} \bx + L_{102} \bx^3 + L_{103} \bx \by^2\right)\partial_{\bx}^2\nn\\
 && + \left(L_{104} \bx + L_{105} \bx^3 + L_{106} \bx \by^2 \right)\partial_{\by}^2\nn\\
&& + \left(L_{107} + L_{108} \bx^2 + L_{109} \by^2\right)\partial_{\bx} + L_{110} \bx \by \partial_{\by},
\label{eq:L1}
\eea
where the explicit forms of the coefficients $\mathcal{A}_{11}, \Omega_{11}, f_{11}, f_{12}$ and $L_{101} - L_{110}$  are specified in the appendix.

\begin{table*}[t!]
\center
\caption{\label{tab:eigenfunctions} Eigenfunctions of the lowest-order modes of a general slender torus with constant specific angular momentum \citep{bla06}. The normalisation constants $a_0-a_5$ are given in table \ref{tab:normalisation} and the coefficients $w_{41}$, $w_{42}$, $w_{51}$, $w_{52}$ are specified in the appendix.}
\begin{indented}
\lineup
\item[]\begin{tabular}{@{}*{7}{l}}
\br
$i$ & \quad $W_i\x{0}$\\
\mr
0   & \quad $a_0$\\
1   & \quad $a_1 \bx$\\
2   & \quad $a_2 \by$\\
3   & \quad $a_3 \bx \by$\\
4   & \quad $a_4\left(1+ w_{41}\bx^2 + w_{42}\by^2 \right)$\\
5   & \quad $a_5\left(1+ w_{51}\bx^2 + w_{52}\by^2 \right)$\\
\br
\end{tabular}
\end{indented}
\end{table*}
%

Substituting (\ref{eq:A1})-(\ref{eq:L1}) into \eref{eq:bij} and considering the eigenmodes derived in \citet{bla06} we find for the radial mode ($i=1$) three non-zero coefficients that correspond to the modes i = 0, 4 and 5 given in table \ref{tab:eigenfunctions}. The resulting eigenfunction $W_{r}$ of the radial epicyclic mode for a slightly non-slender torus in first-order accuracy reads
\be
W_{r}= a_1 \bx + \beta\left(C_0 + C_1 \bx^2 + C_2 \by^2\right) + \mathcal{O} (\beta^2),
\label{eq:W_r}
\ee
where $C_0 = a_0 b_{10} + a_4 b_{14} + a_5 b_{15}$, $C_1 = a_4 b_{14} w_{41} + a_5 b_{15} w_{51}$ and $C_2 = a_4 b_{14} w_{42} + a_5 b_{15} w_{52}$. The normalisation constants $a_0$, $a_4$, $a_5$ are listed in table \ref{tab:normalisation} and the eigenfunction-related terms $w_{41}$, $w_{42}$, $w_{51}$, $w_{52}$ are specified in the appendix. The coefficients $b_{10}$, $b_{14}$, $b_{15}$, given by \eref{eq:bij}, take the following forms
\bea
b_{10} &=& -\frac{a_0 a_1\pi }{4n^2(n+1)(\sigma_0^2-\sigma_1^2)\T^3\R^3}\left\{ 
L_{109}\R^2 + \left[L_{108} + 2n(m -\R)\times\right.\right.\nn\\
&&\left.\left. \times \left\{\mathcal{A}_{11}(m -\R) - 2m\Omega_{11}\right\} + 2L_{107}(n+1)\R^2\right]\T^2 \right\},
\eea
\bea
b_{1q} &=& \frac{a_1 a_q \pi}{8n^2(n+1)(n+2)(\sigma_1^2 - \sigma_q^2)\T^5\R^5}
\left[ \{ L_{108} + 2n\mathcal{A}_{11}(m - \R)^2 \right.\nn\\
&&\left. - 4m^2n\Omega_{11} + 4mn\Omega_{11} \R \} 
 \times \{[2(n+2) \R^2 + 3W_{q1}]\T^2  \right.\nn\\
&&\left. + \R^2 W_{q2}\}\T^2 + L_{109} \{ [2(n+2)\R^2 + W_{q1}]\T^2 + 3\R^2W_{q2} \} \R^2  \right.\nn\\
&&\left. + 2L_{107}(n+2)\T^2\R^2 \{ [2(n+1)\R^2 + W_{q1}]\T^2 + \R^2W_{q2} \} \right],
\eea
where $q=4$ or $q=5$ in case of $b_{14}$ or $b_{15}$, respectively, and $\sigma_0$, $\sigma_1$, $\sigma_4$, $\sigma_5$ are specified in table \ref{tab:eigenfrequencies}.

\begin{table*}[t!]
\centering
\caption{\label{tab:eigenfrequencies} Eigenfrequencies of the eigenmodes given in table \ref{tab:eigenfunctions}.}
\begin{indented}
\lineup
\item[]\begin{tabular}{@{}*{7}{l}}
\br
$i$ \quad & $\sigma_i^2$\\
\mr
0 \quad & 0  \\
1 \quad & $\R^2$\\
2 \quad & $\T^2$\\
3 \quad & $\R^2+\T^2$\\
4 \quad & $\left\{(2n+1)(\R^2+\T^2)-[4n(n+1)
     (\T^2-\R^2)^2+(\R^2+
     \T^2)^2]^{1/2}\right\} / (2n)$\\
5 \quad & $  \left\{(2n+1)(\R^2+\T^2)+[4n(n+1)
    (\T^2-\R^2)^2+(\R^2+
    \T^2)^2]^{1/2}\right\} /(2n)$\\
\br
\end{tabular}
\end{indented}
\end{table*}
%

Similarly, for the vertical epicyclic mode ($i=2$) we find one non-zero coefficient corresponding to $j=3$ and receive the final expression for the eigenfunction $W_{\theta}$ of the vertical epicyclic mode,
\be
W_{\theta}= a_2 \by + \beta C_3 \bx \by + \mathcal{O} (\beta^2)
\label{eq:W_theta}
\ee
with $C_3 = a_3b_{23}$. Again, $a_3$ is the normalisation constant (see table \ref{tab:normalisation}) and the $b_{23}$ coefficient is given by
\bea
b_{23} &=& \frac{a_2a_3\pi}{8n^2(n+2)(n+1)(\sigma_2^2-\sigma_3^2)\T^3\R^3} \left\{L_{110} - 2n(m - \T) \times\right.\nn\\
&&\left.\times \left[\mathcal{A}_{11}(\T -m)  + 2m\Omega_{11}\right]\right\}
\eea
with $\sigma_2$, $\sigma_3$ listed in table \ref{tab:eigenfrequencies}.

\begin{table*}[t!]
\centering
\caption{\label{tab:normalisation} Normalisation constants of the eigenmodes in table \ref{tab:eigenfunctions}  \citep{bla07}. They are calculated such that the eigenfunctions are normalised in the inner product (\ref{eq:inner product}).}
\begin{indented}
\lineup
\item[]\begin{tabular}{@{}*{7}{l}}
\br
$i$ & $a_{i}$\\
\mr
0 \quad & $\left({n \R \T / \pi}\right)^{1/2}$ \\
1 \quad & $a_0[2(n+1)\R^2]^{1/2}$ \\
2 \quad & $a_0[2(n+1)\T^2]^{1/2}$ \\
3 \quad & $a_0[4(n+1)(n+2)\R^2\T^2]^{1/2}$ \\
4 \quad & $a_0\left\{{(n+2)[\sigma_4^2-(\T^2+
     \R^2)] / \left[2n\sigma_4^2-(2n+1)(\T^2
     +\R^2)\right]}\right\}^{1/2}$ \\
5 \quad & $a_0\left\{{(n+2)[\sigma_5^2-(\T^2+
     \R^2)] / \left[2n\sigma_5^2-(2n+1)(\T^2
     +\R^2)\right]}\right\}^{1/2}$ \\
\br
\end{tabular}
\end{indented}
\end{table*}
%

Then, using \eref{eq:A1}-\eref{eq:L1}, we find the terms
\bea
\fl\mathcal{A}\x{1}| W_1\x{0}\rangle = a_1\mathcal{A}_{11}\bx^2, \qquad\qquad & \mathcal{A}\x{1}| W_2\x{0}\rangle = a_2\mathcal{A}_{11}\bx \by, \nn\\
\fl\bar{\Omega}\x{1}|W_1\x{0}\rangle = a_1\Omega_{11}\bx^2, \qquad\qquad & \bar{\Omega}\x{1}|W_2\x{0}\rangle = a_2\Omega_{11}\bx \by, \nn\\
\fl\hat{L}\x{1}|W_1\x{0}\rangle = a_1(L_{107}+L_{108}\bx^2+L_{109}\by^2), \qquad\qquad & \hat{L}\x{1}|W_2\x{0}\rangle = a_2L_{110}\bx \by.
\label{eq:L1W1}
\eea
Substituting \eref{eq:L1W1} into the formula \eref{eq:omega1} for $\omega_i\x{1}$,
we obtain in the inner product for both modes $i=1$ and $i=2$ odd functions of $\bx$ and $\by$, such that the integration over the elliptical torus cross-section yields zero. Therefore, to find the relevant pressure corrections to the radial and vertical mode frequencies, one needs to extend the expansion to second-order in torus thickness.

\subsubsection{Second-order corrections}
\label{sec:second}

The perturbation equation \eref{eq:eigenvalue} expanded to second-order in $\beta$ reads
\bea
\hat{L}\x{0}W_i\x{2}+\hat{L}\x{1}W_i\x{1} + \hat{L}\x{2}W_i\x{0} &=& -2n\left\{\sigma_i^2W_i\x{2} +
(\sigma_i^2\mathcal{A}\x{1} \right.\nn\\
&& \left. - 2m\sigma_i\bar{\Omega}\x{1})W_i\x{1} + \left[\sigma_i^2 \mathcal{A}\x{2} \right.\right.\nn\\
&& \left. \left. + m^2\left(\bar{\Omega}^{(1)}\right)^2 - 2m\sigma_i\bar{\Omega}^{(2)} + 2\sigma_i\bar{\omega}_i\x{2}
 \right.\right.\nn\\  
&&\left.\left. - 2m\sigma_i\mathcal{A}\x{1}\bar{\Omega}^{(1)}\right]W_i\x{0}\right\}.
\label{eq:PP2}
\eea
Analogous to the first-order case one may write
\be
W_i\x{2}=\sum_{j}{c_{ij}W_j\x{0}}
\label{eq:W2}
\ee
and take the inner product of \eref{eq:PP2} with a zeroth-order eigenfunction $W_k\x{0}$. For a subscript $k$ referring to the mode of interest ($k=i$) we obtain the formula for the second-order correction,
\bea
\bar{\omega}_i\x{2} &=& - \frac{1}{4n\sigma_i}\left[ \left\langle W_i\x{0}|\hat{L}\x{2} + 2nm^2\left(\bar{\Omega}^{(1)}\right)^2 - 4nm\sigma_i\bar{\Omega}\x{2}\right.\right.\nn\\
&& \left.\left. + 2n\sigma_i^2\mathcal{A}\x{2} - 4nm\sigma_i\mathcal{A}\x{1}\bar{\Omega}\x{1}|W_i\x{0}\right\rangle\right.\nn\\
&& \left.+ \sum_{j} b_{ij}\left\langle W_i\x{0}|\hat{L}\x{1} + 2n\sigma_i^2\mathcal{A}\x{1} - 4nm\sigma_i\bar{\Omega}\x{1}|W_j\x{0}\right\rangle\right].
\label{eq:omega2}
\eea

The second-order terms of $\mathcal{A}$, $\bar{\Omega}$, $f$ and the $\hat{L}$ operator take now the form
\bea
\mathcal{A}\x{2} &=& \mathcal{A}_{21}\bx^2 + \mathcal{A}_{22}\by^2,\\
\label{eq:A2}
\bar{\Omega}\x{2} &=& \Omega_{21}\bx^2 + \Omega_{22}\by^2,\\
\label{eq:Omega2}
f\x{2} &=& f_{21}\bx^4 + f_{22}\bx^2 \by^2 + f_{23}\by^4,\\
\label{eq:f2}
\hat{L}\x{2} &=& \left\{L_{201}\bx^4 + L_{202}\bx^2 + L_{203}\bx^2 \by^2 + L_{204}\by^2 + L_{205}\by^4\right\}\partial_{\bx}^2\nn\\
&& + \left\{L_{206}\bx^4 + L_{207}\bx^2 + L_{208}\bx^2 \by^2 + L_{204}\by^2 + L_{209}\by^4\right\}\partial_{\by}^2\nn\\
&& + \left\{L_{210}\bx^3 + L_{211}\bx + L_{212}\bx \by^2\right\}\partial_{\bx}  \nn\\
&& + \left\{L_{213}\bx^2\by + L_{214}\by + L_{215}\by^3\right\}\partial_{\by}\nn\\
&&+ \left\{L_{216} + L_{217}\bx^2 + L_{218}\by^2\right\},
\label{eq:L2}
\eea
with the coefficients $\mathcal{A}_{21}$, $\mathcal{A}_{22}$, $\Omega_{21}$, $\Omega_{22}$, $f_{21}$, $f_{22}$, $f_{23}$ and $L_{201}$ - $L_{218}$ again specified in the appendix.

Inserting all first and second-order expansion terms into \eref{eq:omega2} we gain a fairly long expression for $\bar{\omega}_1\x{2}$. Along with \eref{eq:Q} it leads to the resulting formula for the eigenfrequency $\bar{\omega}_r$ of the radial mode. In order to keep the overview we write the expression in the following form
\bea
\bar{\omega}_r &=& \R + m - \frac{\beta^2}{4n\sigma_1} \sum_{l=1}^{4} P_l + \mathcal{O} (\beta^3)
\label{eq:radial}
\eea
where
\bea
P_1 &\equiv& \langle W_1\x{0}|\hat{L}\x{2} + 2nm^2\left(\bar{\Omega}^{(1)}\right)^2 - 4nm\sigma_1\bar{\Omega}\x{2}+ 2n\sigma_1^2\mathcal{A}\x{2}\nn\\
&& -4nm\sigma_1\mathcal{A}\x{1}\bar{\Omega}\x{1}|W_1\x{0}\rangle\nn\\
\bigskip
&=&\frac{a_1^2\pi}{4n(n+1)(n+2)\T^3\R^5}\left\{\left(L_{212} + L_{218}\right)\R^2 \right.\nn\\
&& \left.+ \left[3\left(L_{210} + L_{217} + 2m^2n\Omega_{11}^2\right)\right.\right.\nn\\
&&\left.\left.+ 2\left(L_{211} + L_{216}\right)\left(n+2\right)\R^2\right]\T^2\right.\nn\\
&&\left. - 4mn\left[3(\mathcal{A}_{11}\Omega_{11} + \Omega_{21})\T^2 + \Omega_{22}\R^2\right]\R\right.\nn\\
&&\left. + 2n\left(3\mathcal{A}_{21}\T^2 + \mathcal{A}_{22}\R^2\right)\R^2\right\},
\label{eq:part1}
\eea

\bea
P_2 \, &\equiv& \, b_{10} \, \langle W_1\x{0}|\hat{L}\x{1} + 2n\sigma_1^2\mathcal{A}\x{1} - 4nm\sigma_1^2\bar{\Omega}\x{1}|W_0\x{0}\rangle\nn\\
&=&\frac{a_0 a_1 b_{10}\pi\R\left(\mathcal{A}_{11}\R - 2m\Omega_{11}\right)}{(n+1)\T\R^3},
\label{eq:part2}
\eea

\bea
P_{q-1} \, &\equiv& \, b_{1q} \, \langle W_1\x{0}|\hat{L}\x{1} + 2n\sigma_1^2\mathcal{A}\x{1} - 4nm\sigma_1^2\bar{\Omega}\x{1}|W_q\x{0}\rangle\nn\\
&=& b_{1q} \frac{a_1 a_q \pi}{2n(n+1)(n+2)\T^3\R^5(\T^2-\R^2)} \left\{-2(n+1)\R^2\T^2\times\right.\nn\\
&&\left.\times\left[3\T^2 (L_{102} - L_{105} + L_{108}) + 2(n+2)\T^2\R^2(L_{101} - L_{104}\right.\right.\nn\\
&&\left.\left.  + L_{107}) + \R^2(L_{103} - L_{106} + L_{109} - L_{110})\right] + n\sigma_q^2\left\{\R^4\left[L_{103}\right.\right.\right.\nn\\
&&\left.\left.\left. + L_{109} + 2(2 + n)\T^2(L_{101} + L_{107})\right] - \R^2\T^2\left[L_{106} + L_{110} \right.\right.\right.\nn\\ 
&&\left.\left.\left. + 2(n+2)\T^2 L_{104}\right] + 3\R^2\T^2(L_{102} + L_{108}) - 3 \T^4 L_{105}\right\} + \right.\nn\\ 
&&\left. n(\T^2 - \R^2)\R(\mathcal{A}_{11}\R - 2m\Omega_{11})\left[2(n+2)\R^2\T^2\right.\right.\nn\\
&& \left.\left.+ 3 \T^2 w_{q1} + \R^2 w_{q2}\right]\right\}.
\label{eq:part3+4}
\eea
Here $q=4$ or $q=5$ in case of $P_3$ or $P_4$, respectively.

For the vertical epicyclic mode we find a much shorter expression,
\bea
\bar{\omega}_{\theta} &=&  \T+ m - \beta^2
\frac{a_2^2\pi}{128n^4(n+1)^2(n+2)^2\left(\sigma_2^3 - \sigma_2\sigma_3^2\right)\omega_{\theta}^8\omega_r^6}\times\nn\\
&& \times a_3^2\pi\left[L_{110} + 2n\T\left(\mathcal{A}_{11}\T - 2m\Omega_{11}\right)\right]\left\{3 L_{109}\R^2 + \T^2\left[L_{108}\right.\right.\nn\\ 
&&\left.\left. + L_{110} + 2 L_{107}(n+2)\R^2 + 2n\T(\mathcal{A}_{11}\T - 2m\Omega_{11})\right]\right\}\nn\\ 
&& - 8n^2(n+1)(n+2)\T^3\R^3\left(\sigma_2^2 - \sigma_3^2\right)\left\{3\left(L_{215} + L_{218}\right)\R^2 + \T^2\left[L_{213}\right.\right.\nn\\
&&\left.\left. + L_{217} + 2m^2n\Omega_{11}^2 + 2\left(L_{214} + L_{216}\right)(n+2)\R^2\right] - 4mn\left[\left(\mathcal{A}_{11}\Omega_{11}\right.\right.\right.\nn\\
&&\left.\left.\left. + \Omega_{21}\right)\T^2 + 3\Omega_{22}\R^2\right]\T + 2n\left(\mathcal{A}_{21}\T^2 + 3\mathcal{A}_{22}\R^2\right)\T^2\right\}\nn\\
&& + \mathcal{O}(\beta^3).
\label{eq:vertical}
\eea

\section{The properties of epicyclic modes for non-slender tori}
\label{sec:figures}

The equations \eref{eq:W_r}, \eref{eq:W_theta}, \eref{eq:radial} and \eref{eq:vertical} represent the epicyclic mode eigenfunctions and eigenfrequencies of non-slender tori as functions of $r_0$, $a$, $m$, $n$ and $\beta$. In this section we use these formulas to illustrate the behaviour for the axisymmetric ($m=0$) and lowest-order non-axisymmetric ($m=\pm 1$) epicyclic modes for variable torus thickness and varying black hole spin value. We take the polytropic index $n=3$ that refers to a radiation-pressure dominated torus \footnote{Note that varying $n$ makes no relevant difference to the results.}.

All figures that are related to frequency behaviour (\ref{fig:radial}-\ref{fig:rapid-axi}, \ref{fig:radial-plus}-\ref{fig:rapid-nonaxi}) display the frequencies defined as $\nu=\omega/2\pi$ \footnote{As it is commonly used, throughout the paper we refer to quantities $\omega$ as to 'frequencies' as well, although in exact notation they should be called 'angular velocities'.}. The horizontal axis in all these figures starts at the radius of the marginally stable orbit calculated for the appropriate black hole spin.

\subsection{Axisymmetric epicyclic modes}
\label{sec:axi}

\begin{center}
\begin{figure}
\begin{center}
\includegraphics[width=.48\hsize]{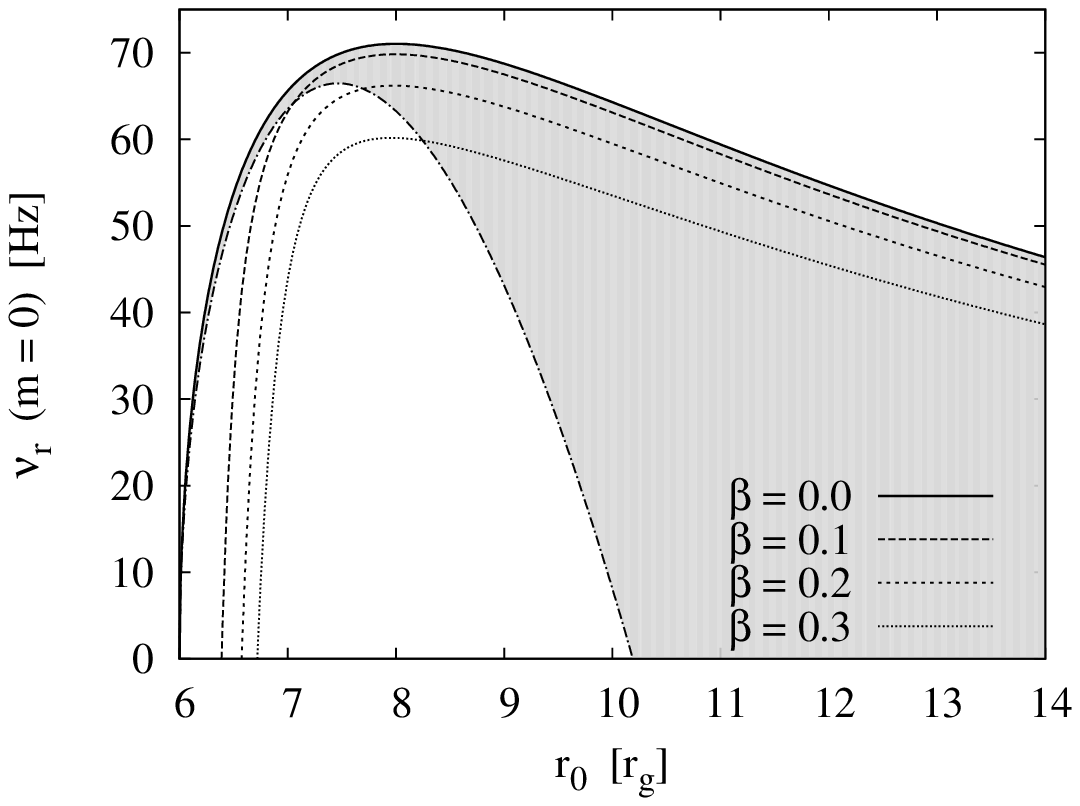} 
\hfill
\includegraphics[width=.48\hsize]{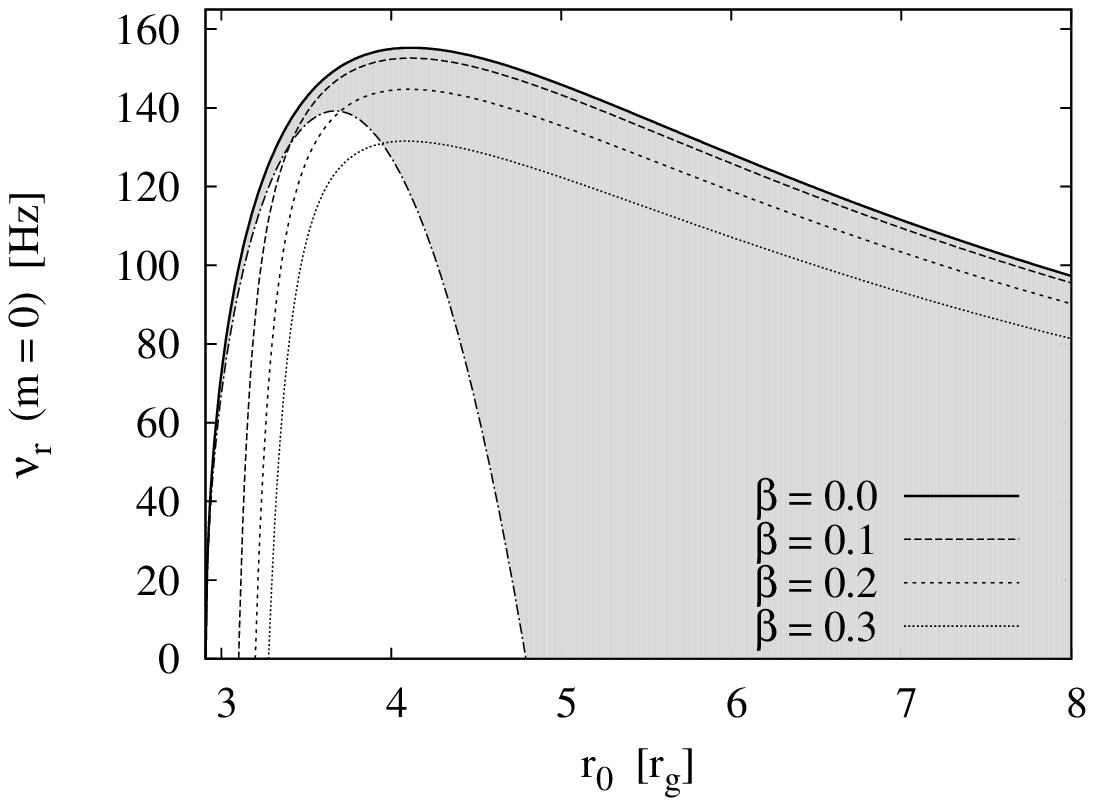}
\end{center}
\caption{\label{fig:radial} The axisymmetric radial epicyclic frequency as a function of the torus pressure maximum $r_0$ for tori of various thickness that orbit a black hole of mass $M=10 M_{_{\odot}}$. The allowed frequencies corresponding to equilibrium tori lie inside the shaded region. {\it Left\/} In the Schwarzschild case $a=0$. {\it Right\/} For a rotating black hole of $a=0.8$.}
\end{figure}
\end{center}

\begin{center}
\begin{figure}
\begin{center}
\includegraphics[width=.48\hsize]{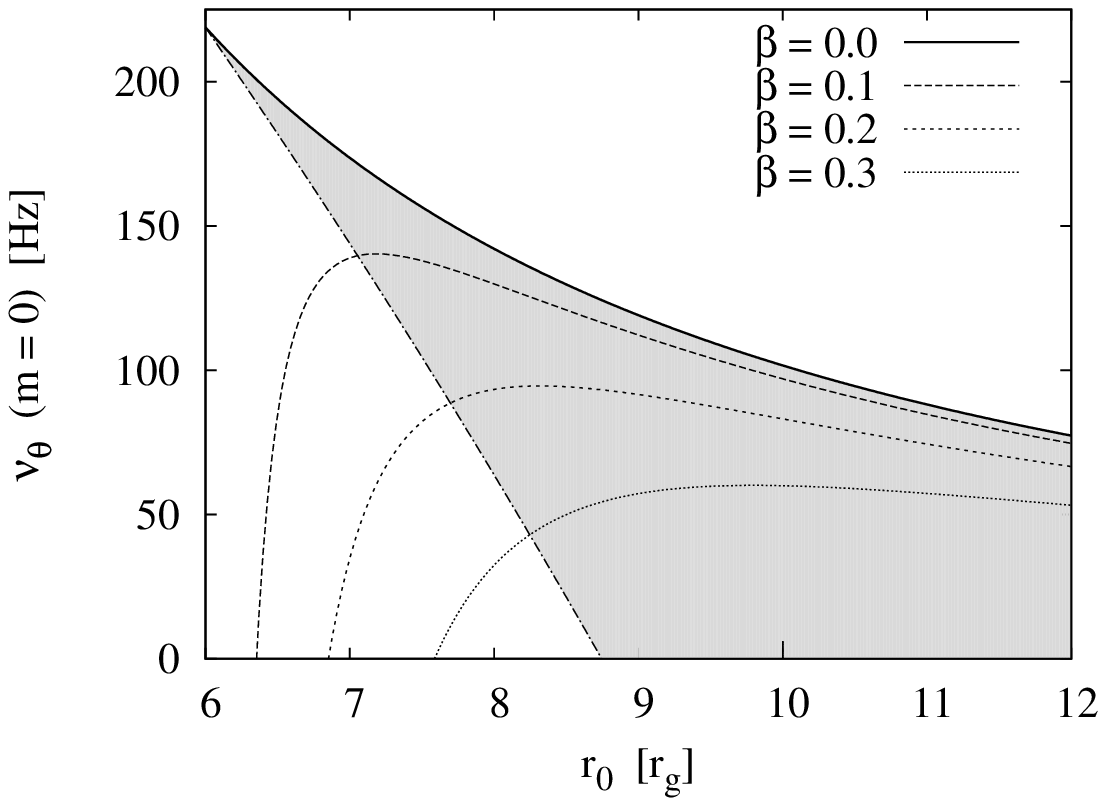}
\hfill
\includegraphics[width=.48\hsize]{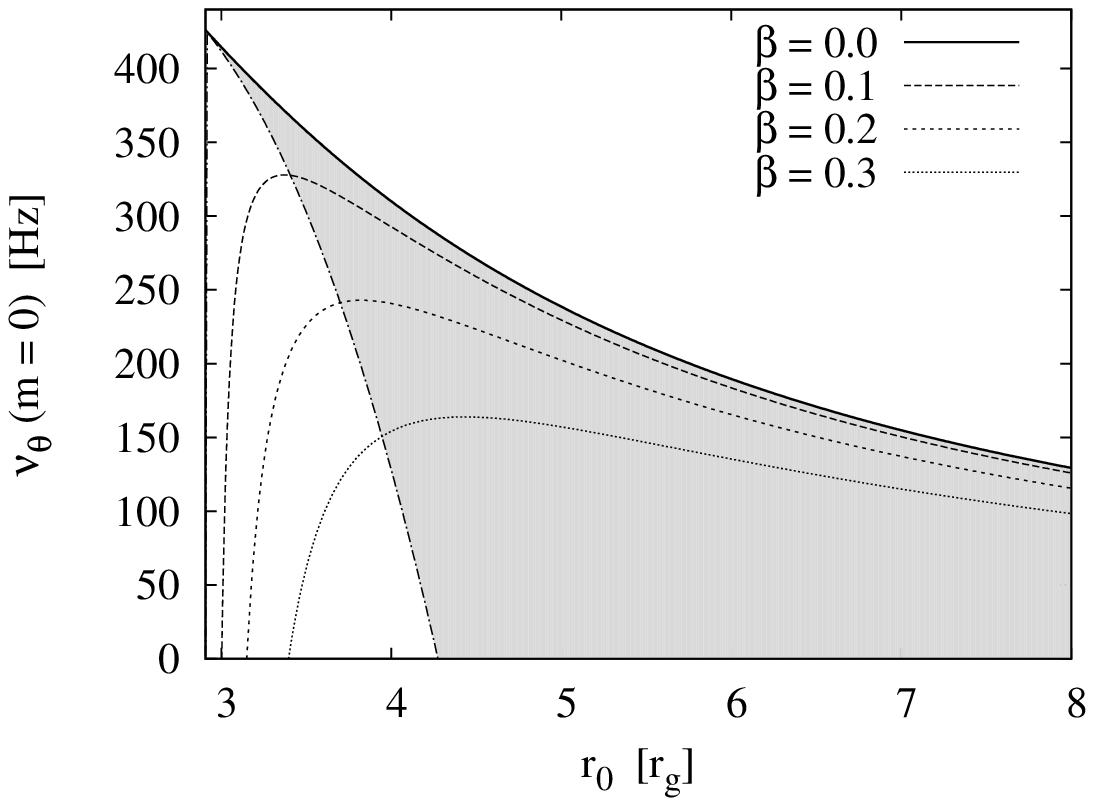}
\end{center}
\caption{\label{fig:vertical} The same as figure \ref{fig:radial}, but for the axisymmetric vertical epicyclic mode.}
\end{figure}
\end{center}

\begin{center}
\begin{figure}
\begin{center}
\includegraphics[width=.48\hsize]{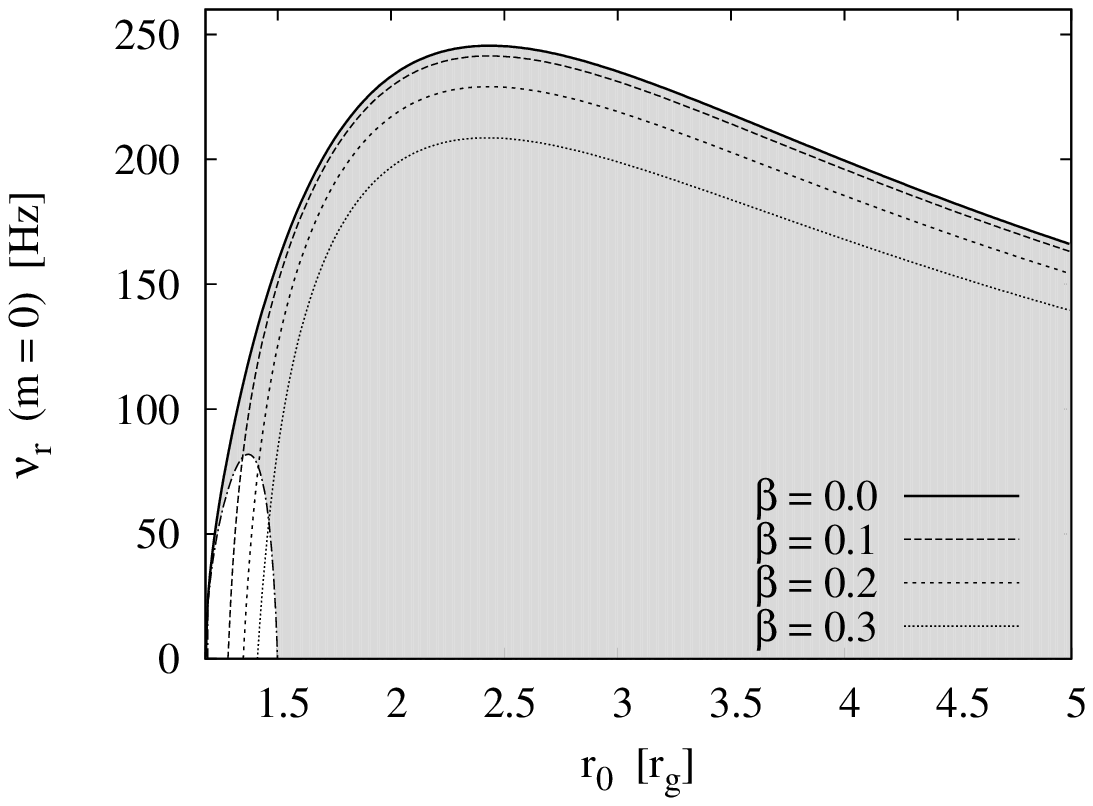}
\hfill
\includegraphics[width=.48\hsize]{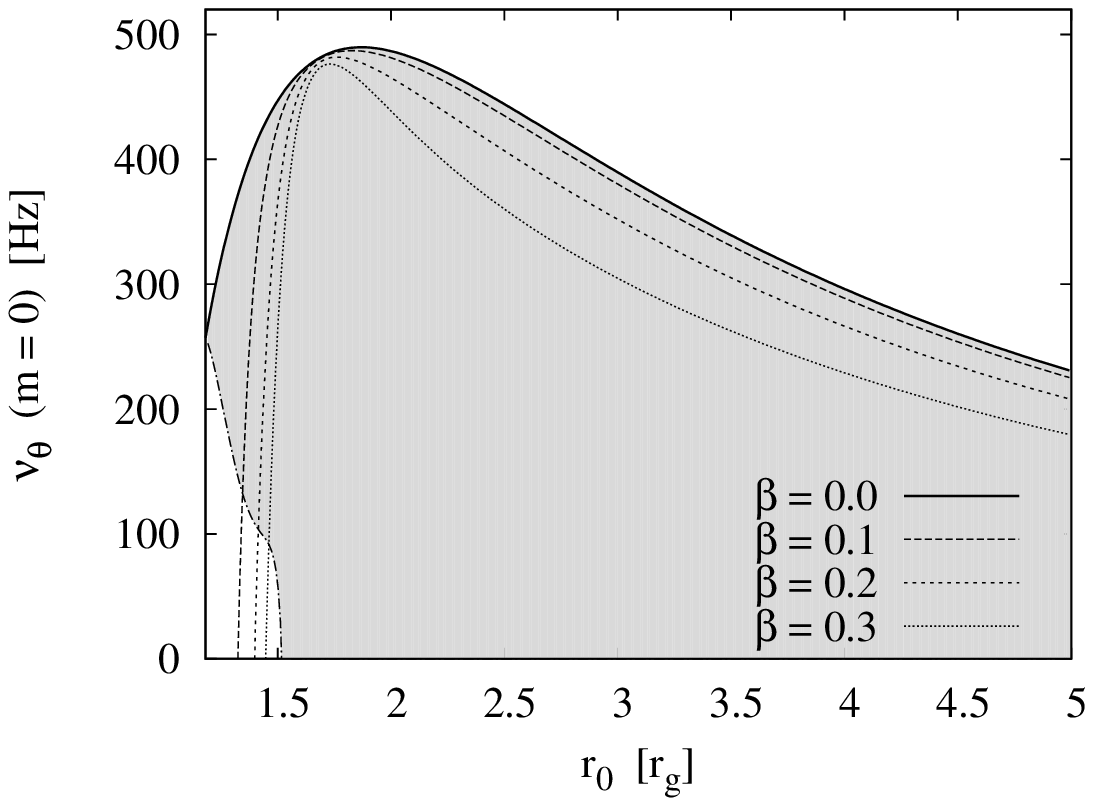}
\end{center}
\caption{\label{fig:rapid-axi} The same as figure \ref{fig:radial}, but for the axisymmetric radial ({\it left\/}) and vertical ({\it right\/}) epicyclic mode in case of a near-extreme Kerr black hole of $a=0.999$.}
\end{figure}
\end{center}

In the axisymmetric ($m=0$) case the $\phi$-components vanish from the equations and the problem becomes symmetric with respect to the rotational axis ($\theta=0$). The corresponding axisymmetric radial and vertical epicyclic frequencies are shown in figures \ref{fig:radial}, \ref{fig:vertical} and \ref{fig:rapid-axi}. The $\beta=0$ line always refers to the epicyclic frequency of a test particle, while the $\beta > 0$ lines illustrate the behaviour of the two frequencies when the torus becomes thicker. As a result of the topology of the equipotential surfaces there is an upper limit on $\beta$ for a given torus pressure maximum above which no closed equipotential surfaces and consequently no equilibrium tori may exist. This limit is incorporated into the figures as a dash-dotted line that defines the region of 'allowed frequencies' above it and to its right (inside the shaded region).

For any black hole spin both axisymmetric frequencies decrease with increasing torus thickness. A comparison  of the left ($a=0$) and right ($a=0.8$) panels of figures  \ref{fig:radial} and \ref{fig:vertical} and the respective panel of \fref{fig:rapid-axi} ($a=0.999$) illustrates the influence of the black hole rotation. 

The radial frequency qualitatively retains for all spin values the same profile albeit the torus thickness. The radius where it becomes zero moves, however, away from the central object. It is interesting to note that for all values of $a$ the frequency maximum for a fixed $\beta$ is reduced by almost exactly the same relative amount\footnote {This does not apply to the {\it actual\/} thickness of the torus, since a given $\beta$ implies for different $a$ a different extent of the torus.}. Moreover, for a non-extremely rotating black hole with $a \lesssim 0.96$, this is more or less true for any location of the pressure maximum $r_0$, while for $a \gtrsim 0.96$ the frequencies at small radii tend to crowd together, as may be seen in figure \ref{fig:rapid-axi}.

The vertical frequency for a Schwarzschild black hole changes its profile with increasing torus thickness from a monotonic function in the case of test particle frequency to a function exhibiting a maximum value (see the left panel of \fref{fig:vertical}). For a Kerr black hole the non-monotonicity is already present for the test particle frequency (although for low to moderate spin values the frequency maximum is located at radii inside the marginally stable orbit) and it remains also for a thicker torus (see the right panels of figures \ref{fig:vertical} and \ref{fig:rapid-axi}). At very high spin values ($a \gtrsim 0.96$), the frequency shape is modified in such a way that the frequency maxima for all tori almost coincide with the $\beta=0$ curve (figure \ref{fig:rapid-axi}, right panel).

\begin{center}
\begin{figure}
\begin{center}
\includegraphics[height=.48\textwidth,angle=-90]{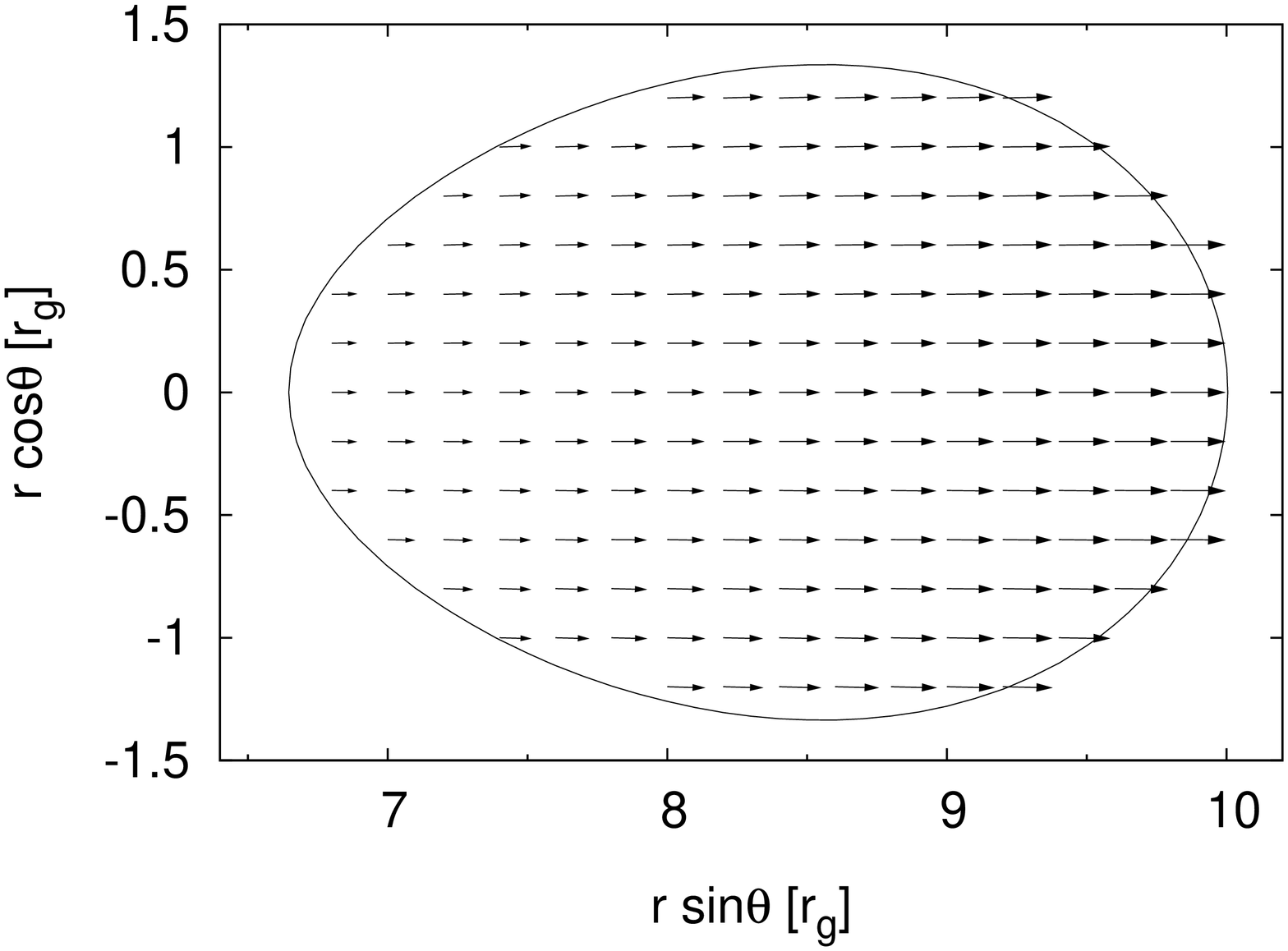} 
\hfill
\includegraphics[height=.48\textwidth,angle=-90]{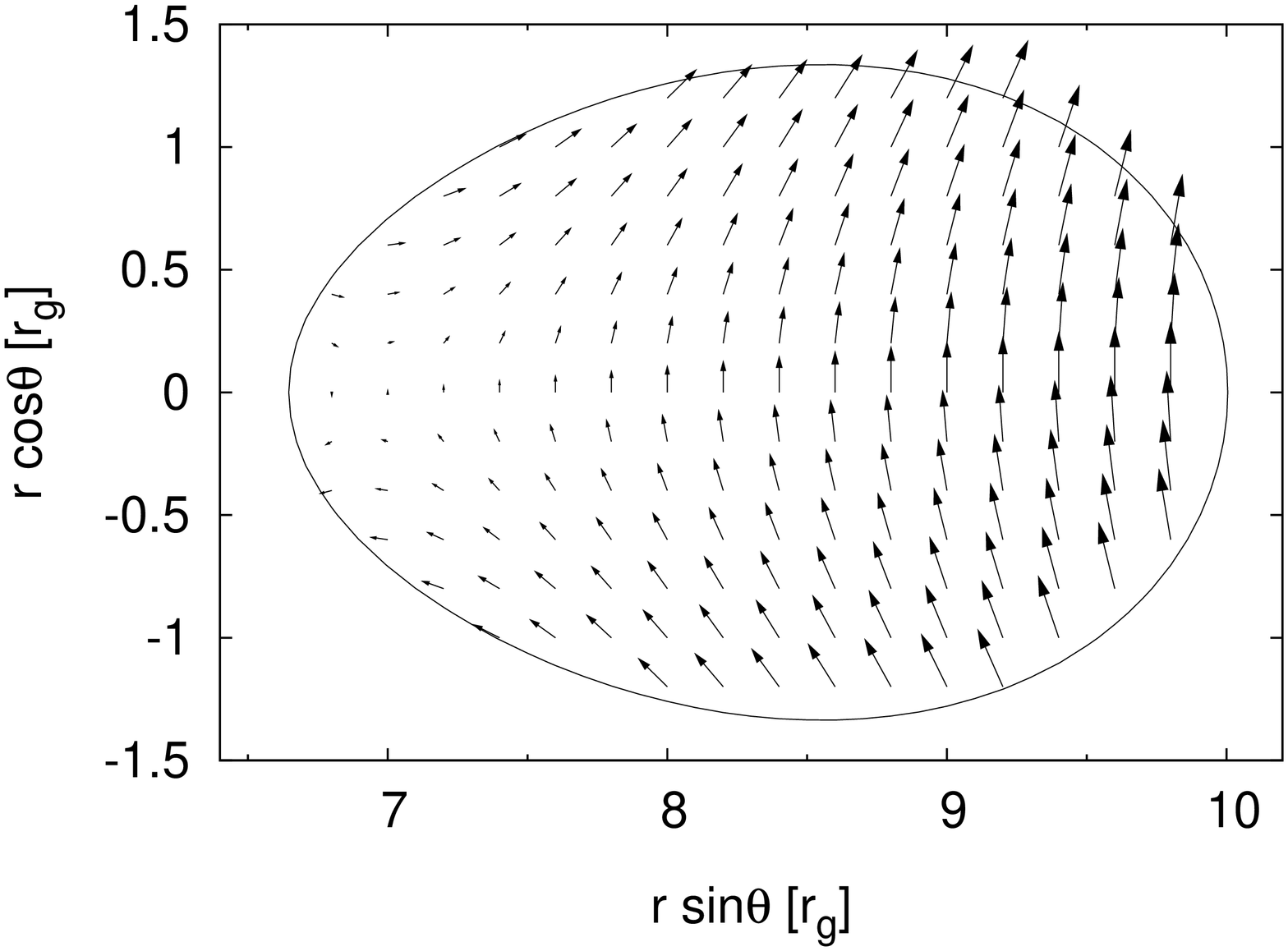}
\end{center}
\caption{\label{fig:velocity-axi} Poloidal flow velocity fields of the axisymmetric radial ({\it left\/}) and vertical ({\it right\/}) epicyclic mode for a torus of $\beta=0.15$ with pressure maximum at $r_0=8 r_{\rm{g}}$ orbiting a Kerr black hole with $a=0.5$.}
\end{figure}
\end{center}

In order to illustrate the characteristic features of the flow for the modes of a thicker torus we use equations \eref{eq:velocity}, \eref{eq:W_r} and \eref{eq:W_theta} to plot the corresponding poloidal velocity fields. The axisymmetric radial mode shows a relatively coherent flow with only slight deviations from the radial motion in the outer regions of the torus (see the left panel in \fref{fig:velocity-axi}). The axisymmetric vertical mode, however, shows a more complex behaviour. As seen in the right panel of \fref{fig:velocity-axi}, it involves vertical motions near the inner and outer edges of the torus, as well as smaller radial flows in regions close to the pressure maximum. Its velocity pattern exhibits the features as one of the lowest-order slender torus modes calculated in \citet{bla06}, the so-called x-mode (see their figure 1). The velocity fields of both modes keep their characteristics similar to those described above independently of the black hole spin. 

\subsection{Non-axisymmetric epicyclic modes}
\label{sec:non-axi}

\begin{center}
\begin{figure}
\begin{center}
\includegraphics[width=.48\textwidth]{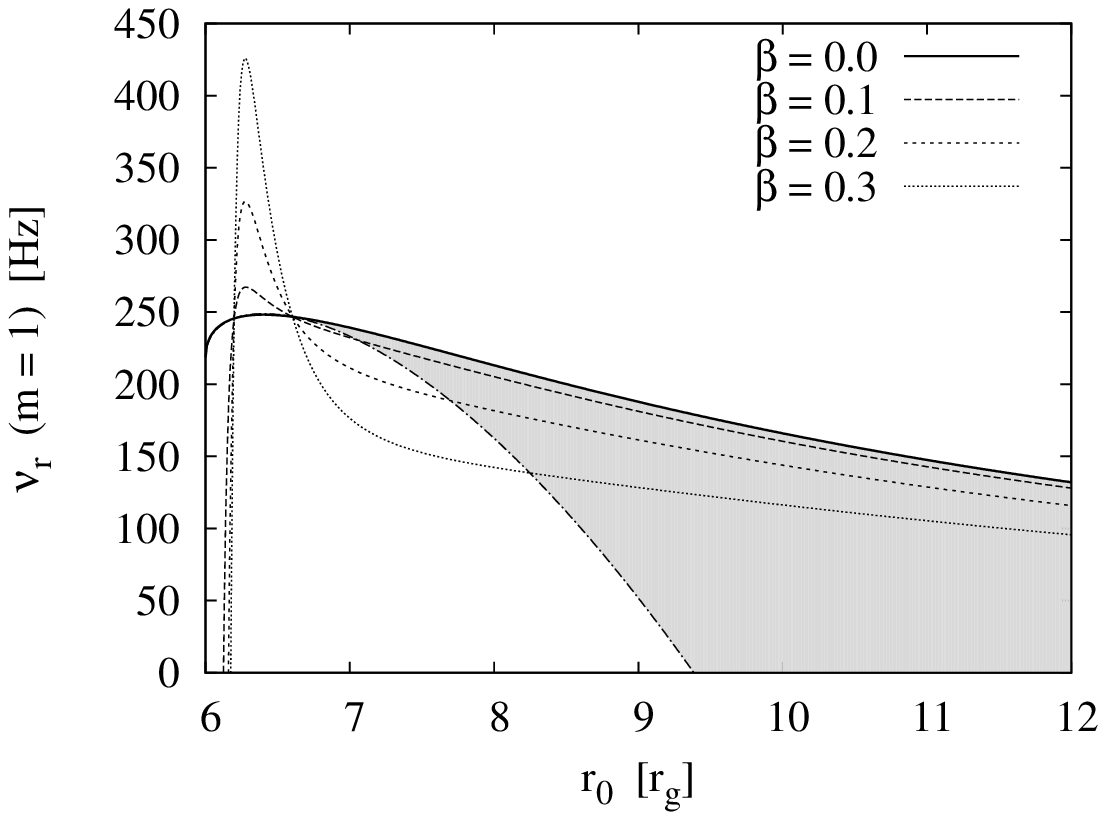}
\hfill
\includegraphics[width=.48\textwidth]{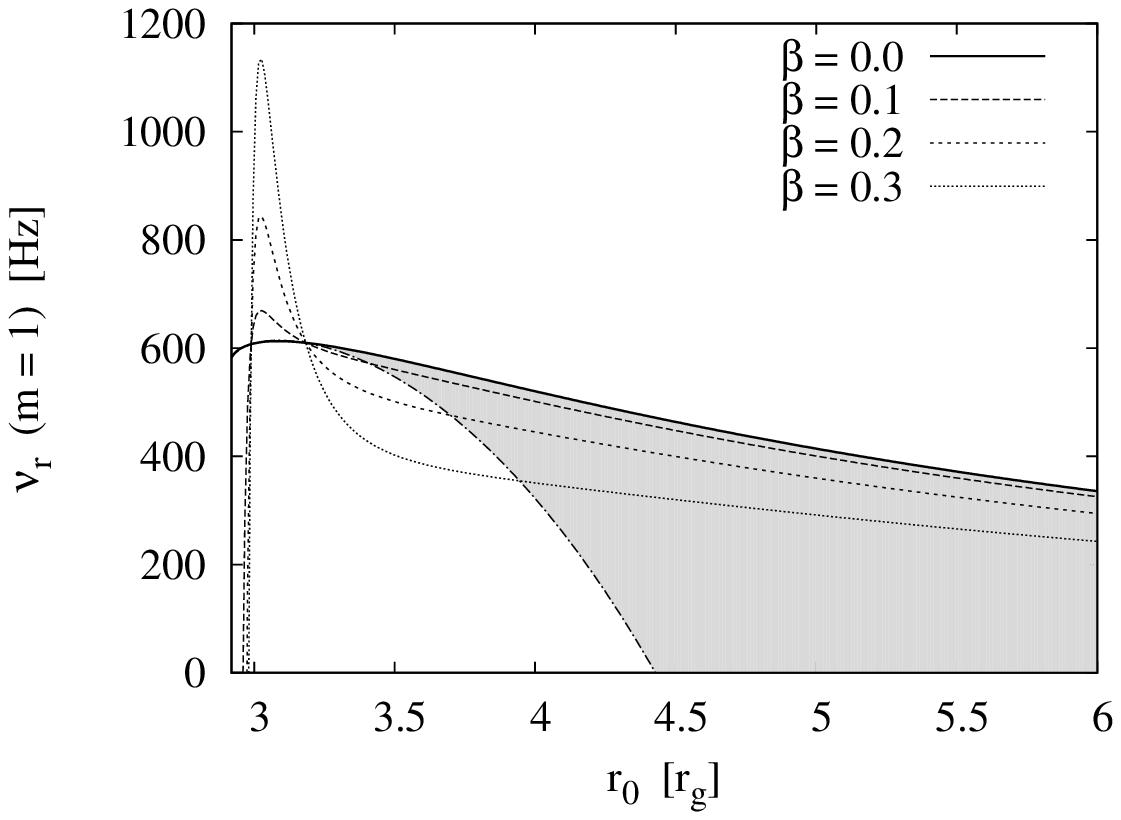}
\bigskip

\includegraphics[width=.48\textwidth]{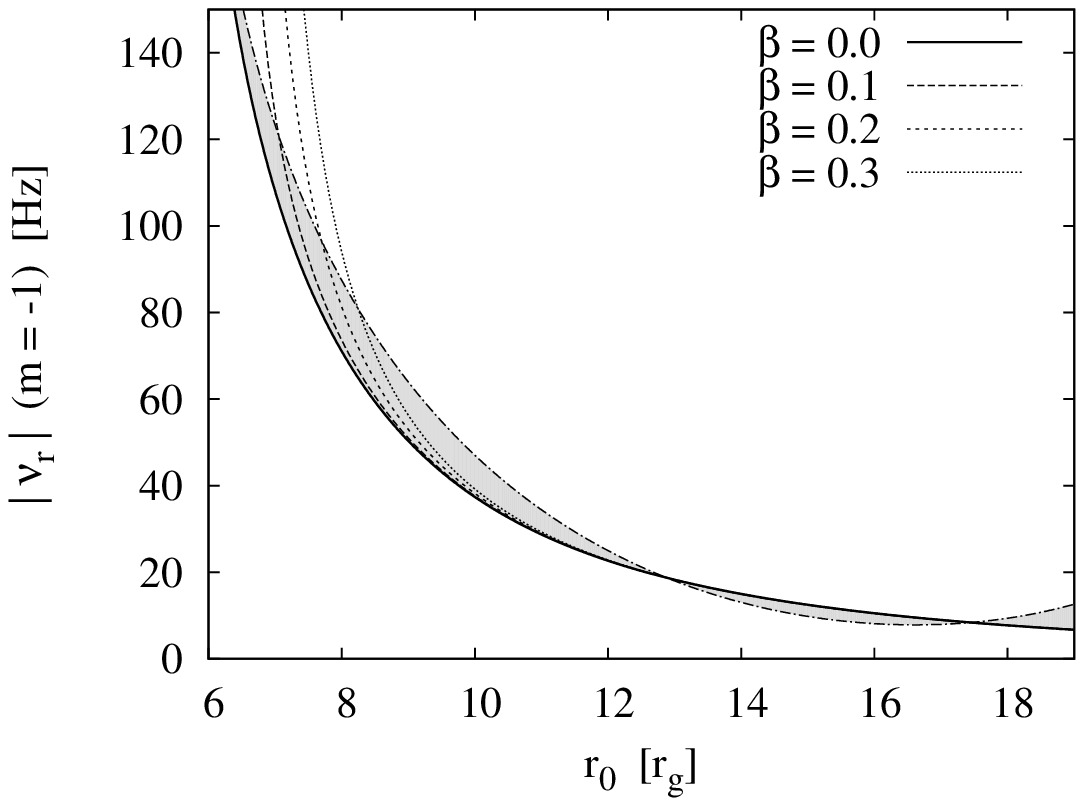}
\hfill
\includegraphics[width=.48\textwidth]{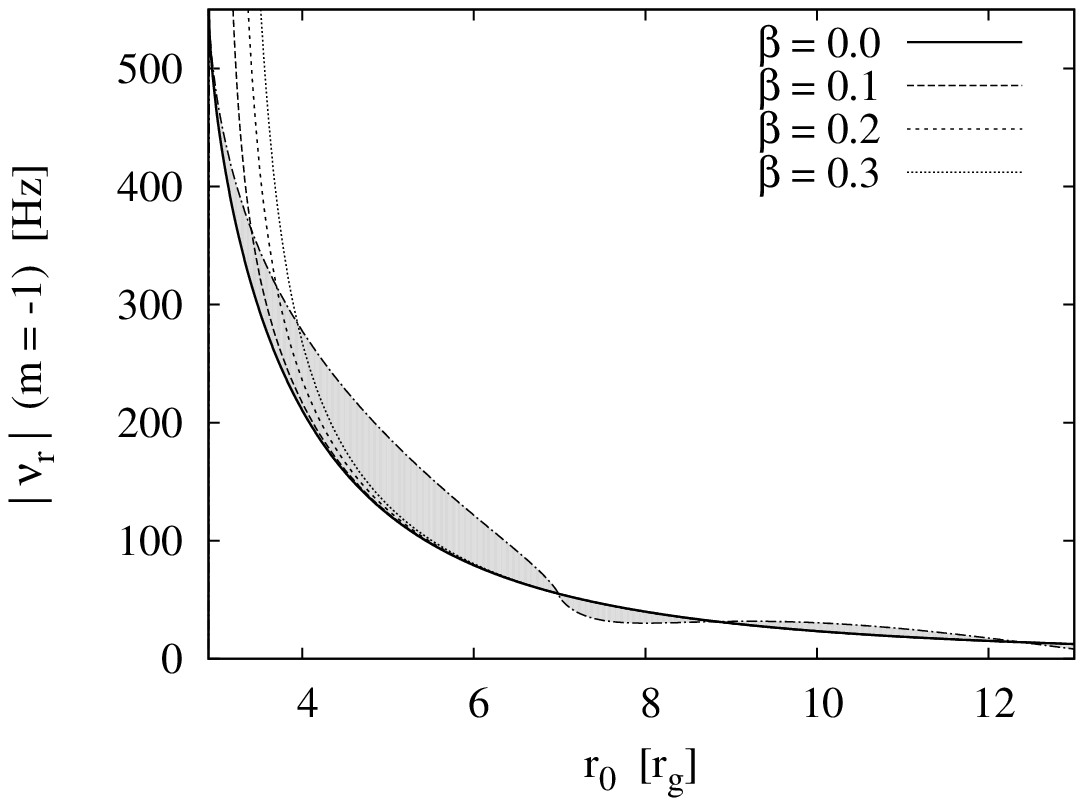}
\end{center}
\caption{\label{fig:radial-plus} \label{fig:radial-minus} The non-axisymmetric $m=1$ ({\it top \/}) and $m=-1$ ({\it bottom\/}) radial epicyclic frequency as a function of $r_0$ for different values of $\beta$ for a $M=10 M_{_{\odot}}$ black hole. {\it Left\/} For a non-rotating black hole. {\it Right\/} For a spinning black hole of $a=0.8$.}
\end{figure}
\end{center}

\begin{center}
\begin{figure}
\begin{center}
\includegraphics[width=.48\textwidth]{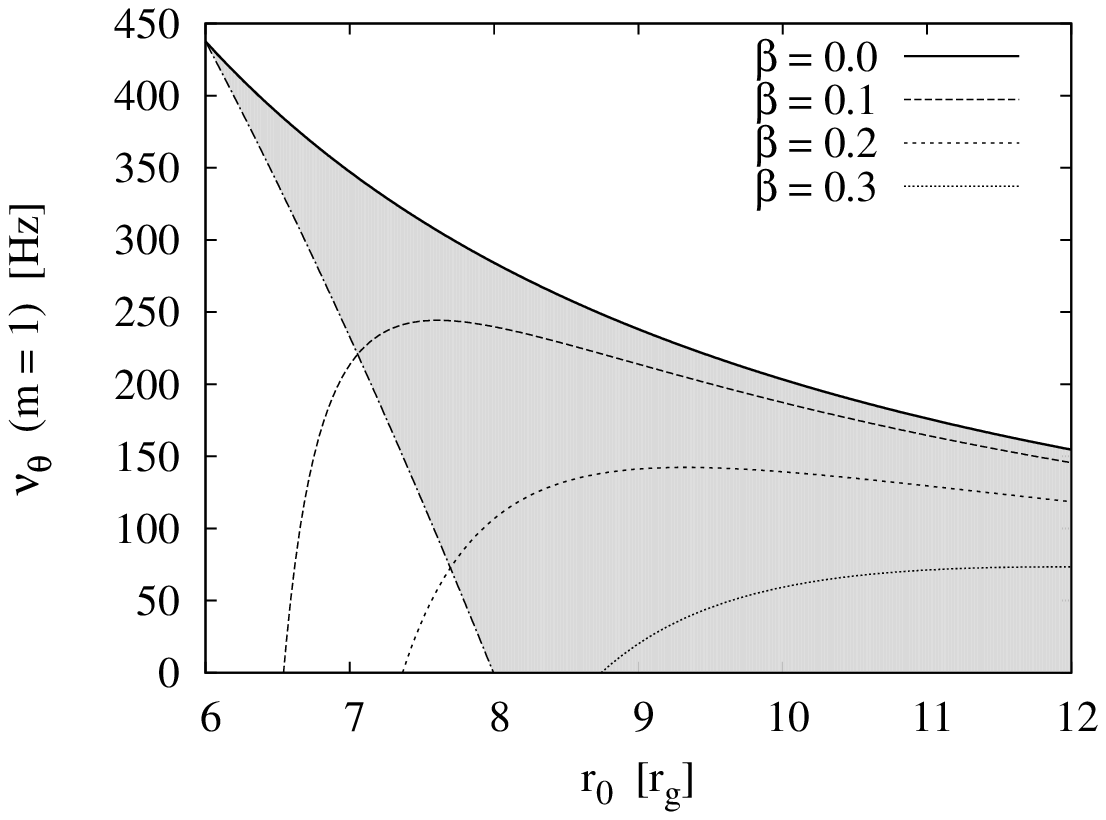}
\hfill
\includegraphics[width=.48\textwidth]{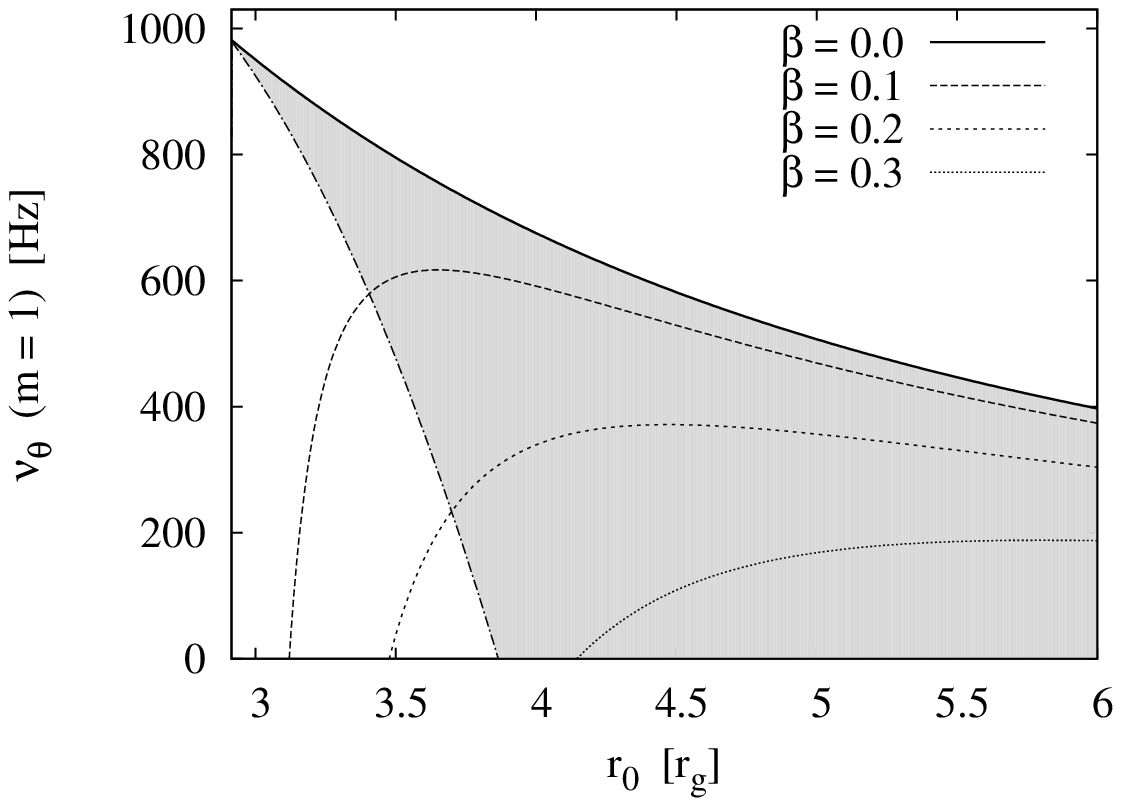}
\bigskip

\includegraphics[width=.48\textwidth]{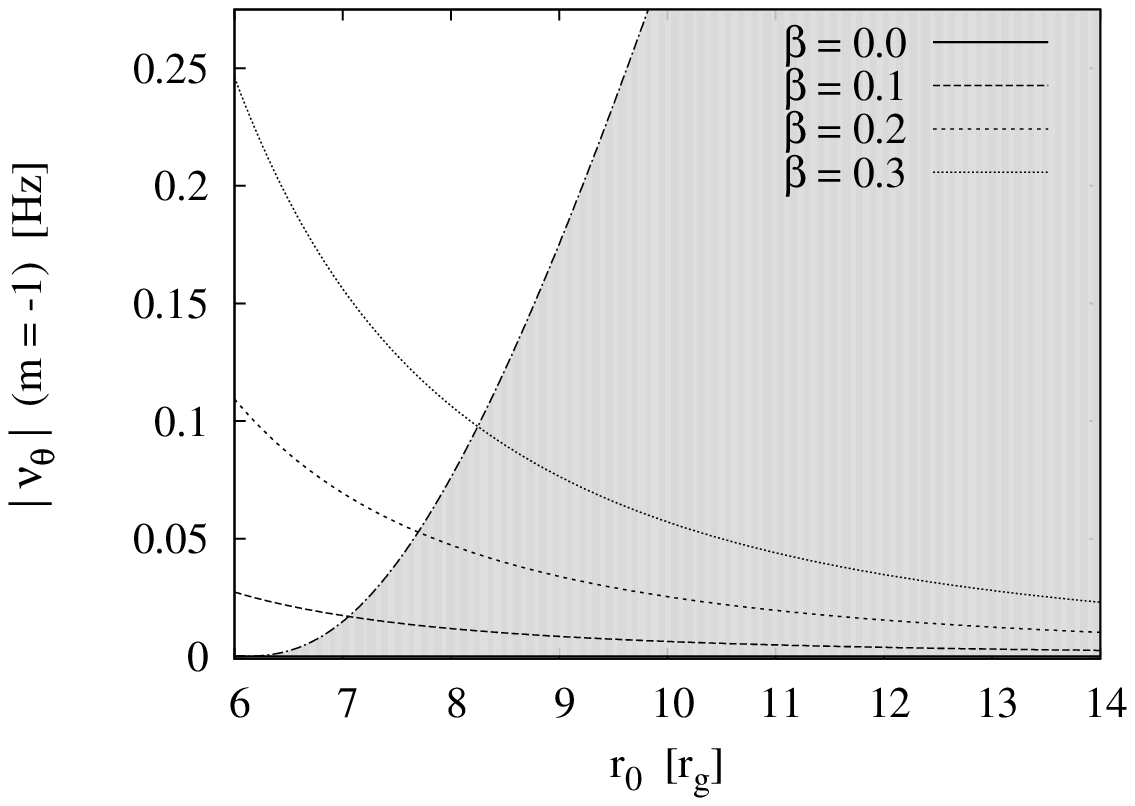}
\hfill
\includegraphics[width=.48\textwidth]{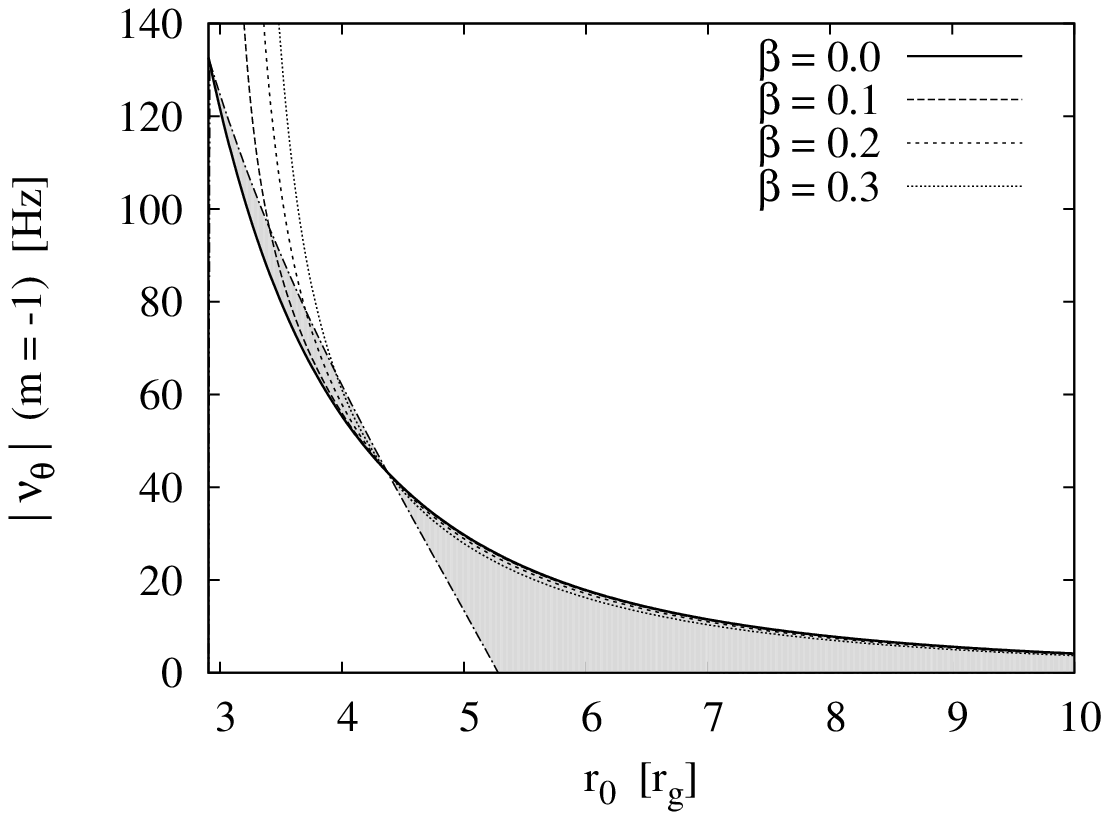}
\end{center}
\caption{\label{fig:vertical-plus} \label{fig:vertical-minus} 
The same as figure \ref{fig:radial-plus}, but for the vertical epicyclic mode.}
\end{figure}
\end{center}

\begin{center}
\begin{figure}
\begin{center}
\includegraphics[width=.48\textwidth]{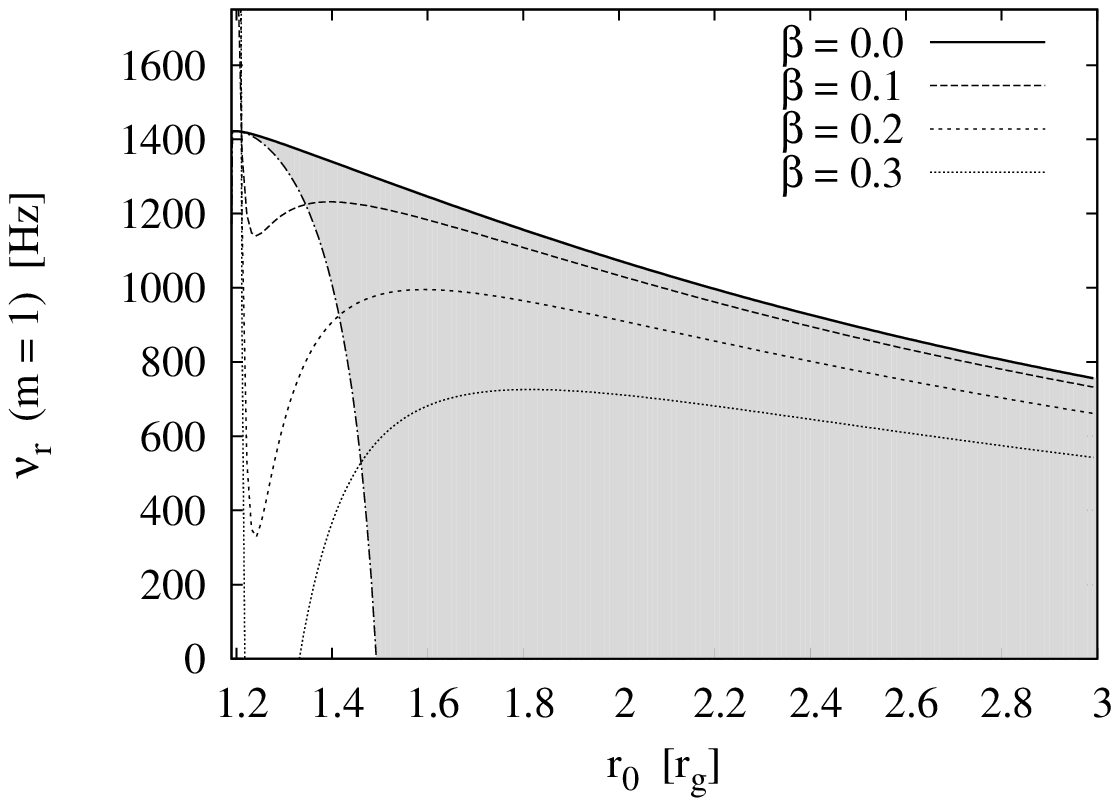}
\hfill
\includegraphics[width=.48\textwidth]{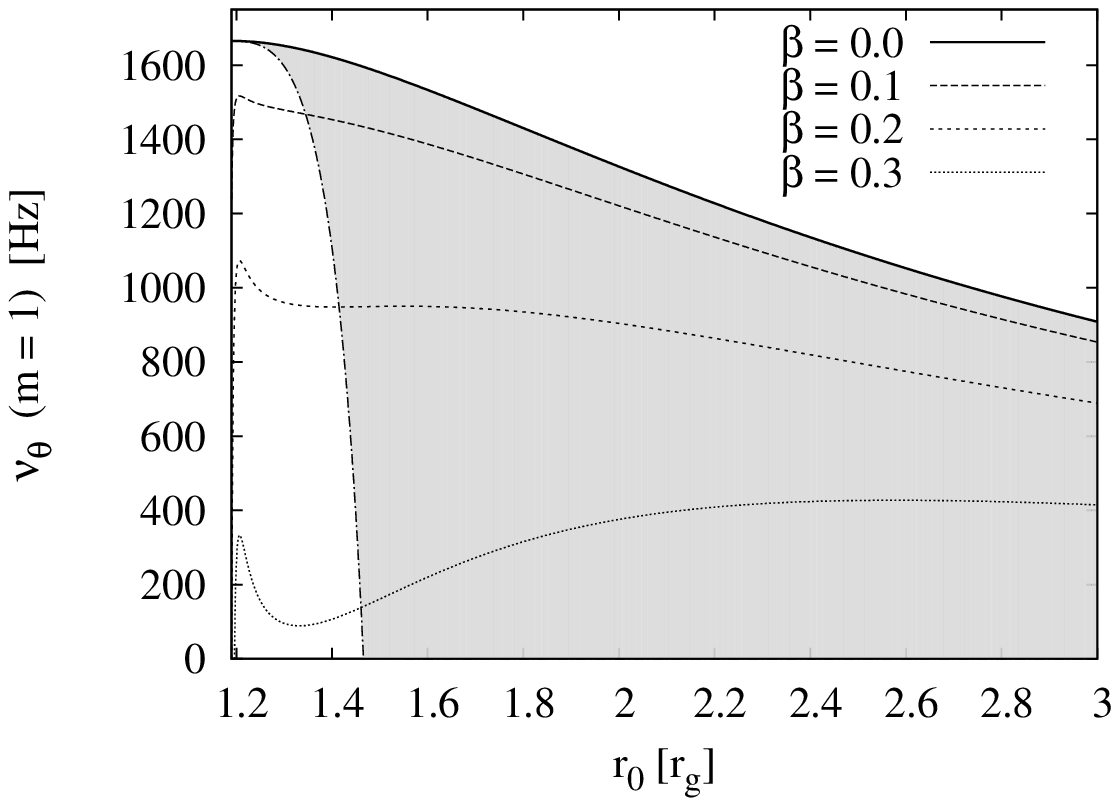}
\bigskip

\includegraphics[width=.48\textwidth]{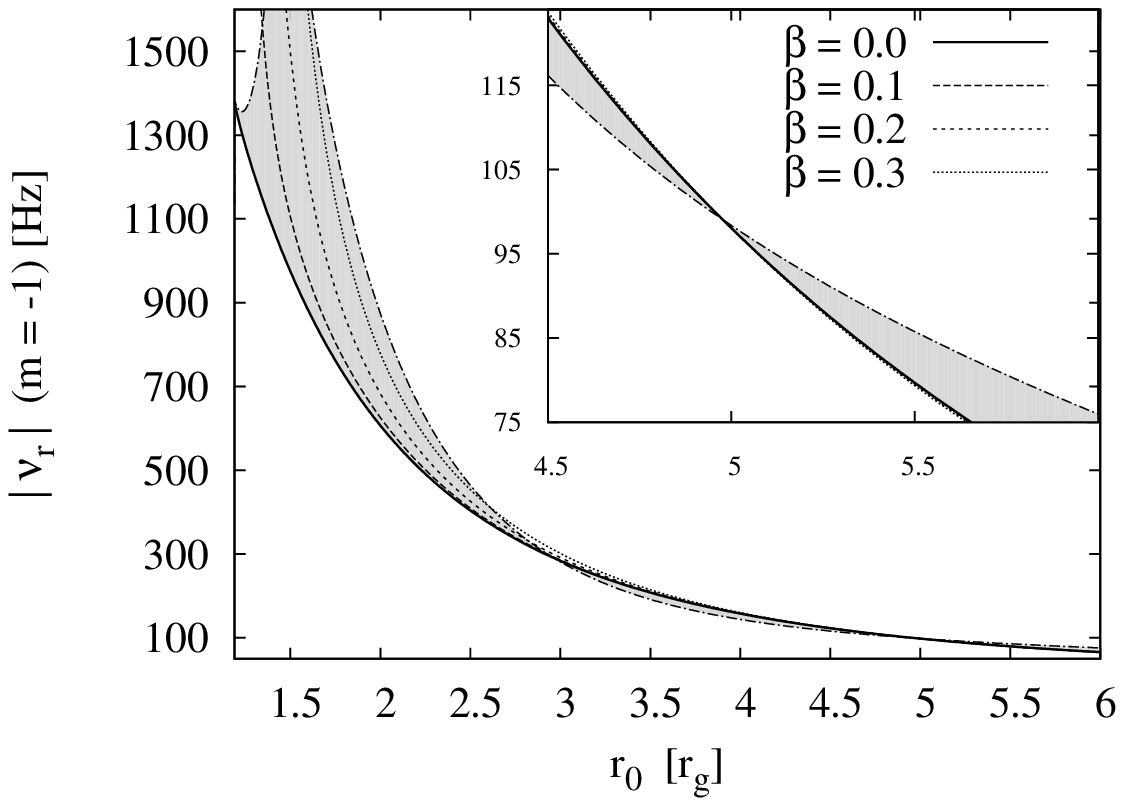}
\hfill
\includegraphics[width=.48\textwidth]{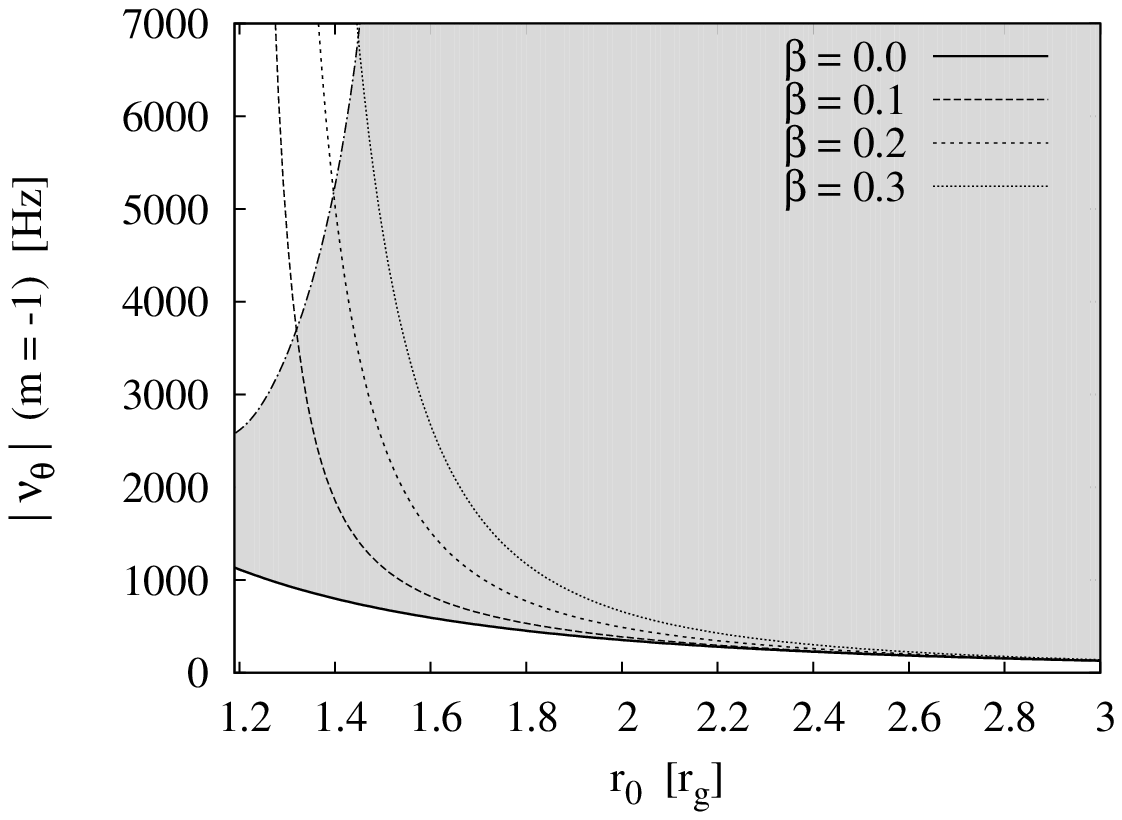}
\end{center}
\caption{\label{fig:rapid-nonaxi-m1}\label{fig:rapid-nonaxi} The non-axisymmetric $m=1$ ({\it top\/}) and $m=-1$ ({\it bottom\/}) epicyclic mode frequencies in case of a near-extreme Kerr black hole of $a=0.999$.
{\it Left\/} The radial epicyclic mode. {\it Right\/} The vertical epicyclic mode.}
\end{figure}
\end{center}

The lowest-order non-axisymmetric ($m=\pm 1$) radial and vertical epicyclic mode frequencies are shown in figures \ref{fig:radial-plus}-\ref{fig:rapid-nonaxi}. Again, the $\beta=0$ curve in each case refers to the test particle frequency and the dash-dotted line specifies the region of allowed frequencies (inside the shaded region). Like above, the left panels of figures \ref{fig:radial-plus} and \ref{fig:vertical-minus} display the frequencies for tori that orbit a Schwarzschild black hole ($a=0$), while the corresponding right panels show the same for a rapidly spinning Kerr black hole ($a=0.8$). Figure \ref{fig:rapid-nonaxi-m1} then displays the frequencies for a near-extreme Kerr black hole ($a=0.999$).

The $m=1$ radial epicyclic mode frequency of a slightly non-slender torus is given by equation \eref{eq:radial}. For a test particle ($\beta=0$) it reduces to the sum of $\Omega_0$ and the test particle axisymmetric radial epicyclic frequency $\omega_{r0}$. For all values of $a$ this frequency decreases with increasing torus thickness (see the top panels of \fref{fig:radial-plus} and the top left panel of \fref{fig:rapid-nonaxi-m1}).

The $m=-1$ radial mode frequency, again given by \eref{eq:radial}, is in the test particle case represented (in absolute value) by the difference between $\Omega_0$ and $\omega_{r0}$. At small radii the frequencies for all spin values increase with growing torus thickness (see the bottom panels of \fref{fig:radial-minus} and the bottom left panel of \fref{fig:rapid-nonaxi}). With rising $r_0$, the non-slender torus frequencies start oscillating about the $\beta=0$ frequency and eventually converge to the test particle profile.

In spherically symmetric spacetimes, the orbital frequency $\Omega_0$ and the axisymmetric vertical epicyclic frequency $\omega_{\theta0}$ of  a test particle coincide. This is no longer true in case of an axially symmetric (Kerr) spacetime or the frequencies of a non-slender torus. The non-axisymmetric $m=1$ vertical epicyclic frequency (given by \eref{eq:vertical}) for test particles corresponds to the sum of $\Omega_0$ and $\omega_{\theta0}$. For a non-slender torus the frequency behaves up to $a \lesssim 0.96$ similarly to the axisymmetric $\omega_\theta$ (compare the top panels of \fref{fig:vertical-plus} to \fref{fig:vertical}). Its form slightly differs only for very high spin values ($a \gtrsim 0.96$) (see the top right panel of \fref{fig:rapid-nonaxi-m1}). 

The $m=-1$ vertical mode frequency (equation \eref{eq:vertical}) is for a test particle given (in absolute value) by the difference between $\Omega_0$ and $\omega_{\theta0}$. In a Schwarzschild spacetime this difference equals zero for test particles, but for increased torus thickness the frequency grows (bottom left panel of \fref{fig:vertical-minus}). For a Kerr black hole, the frequency converges for all spin values to the test particle frequency as the pressure maximum $r_0$ moves away from the black hole. At smaller radii and for $a \lesssim 0.96$ there is a crossing point with the $\beta=0$ frequency, such that, in the first interval, rising $\beta$ causes the frequency to increase in contrast to the second interval where the frequencies show opposite behaviour (\fref{fig:vertical-minus}, bottom right panel). For $a \gtrsim 0.96$ the crossing point vanishes and rising torus thickness evokes rising frequencies at all radii (bottom right panel of \fref{fig:rapid-nonaxi-m1}).

\begin{center}
\begin{figure}
\begin{center}
\includegraphics[height=.48\textwidth,angle=-90]{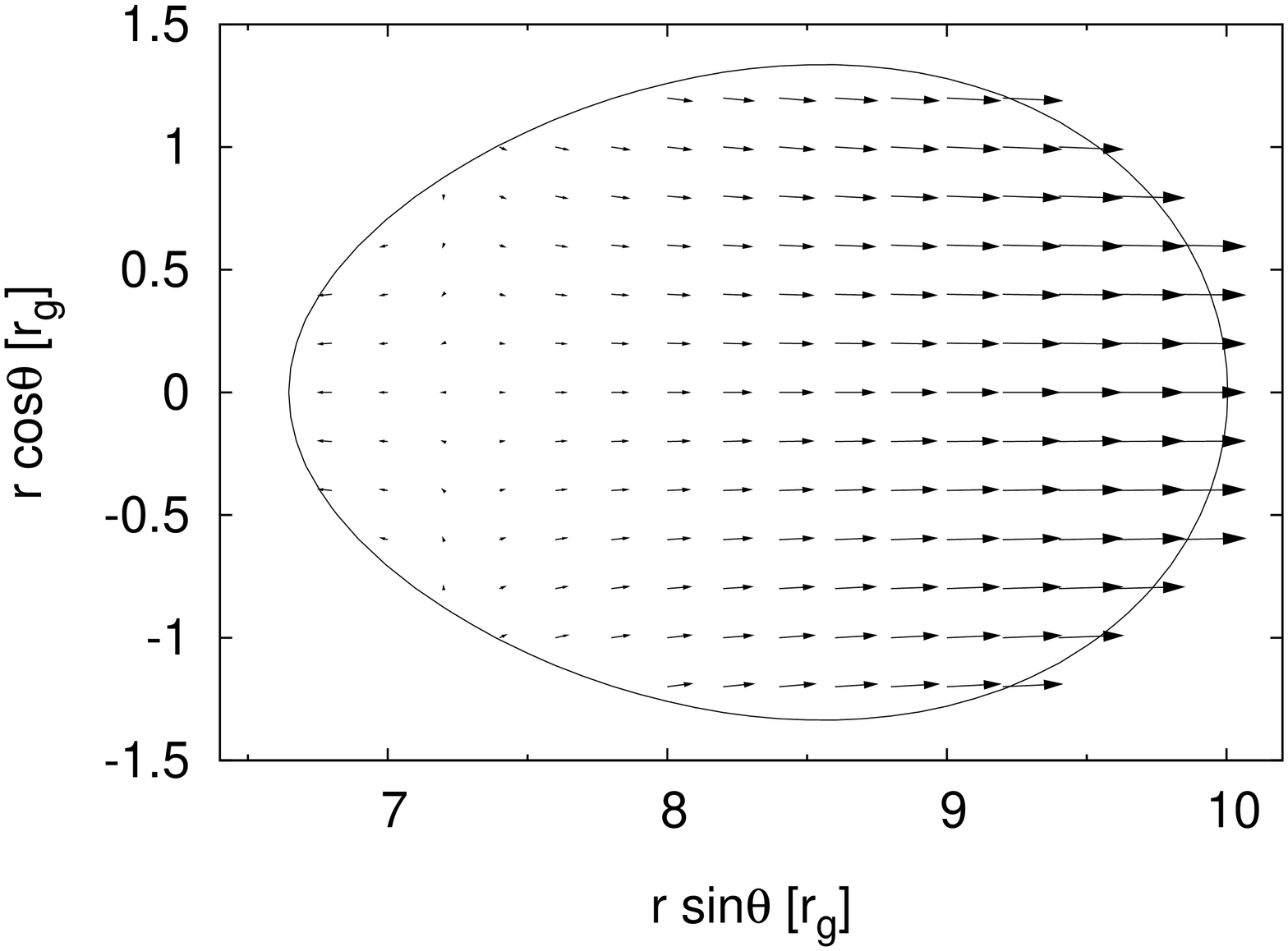} 
\hfill
\includegraphics[height=.48\textwidth,angle=-90]{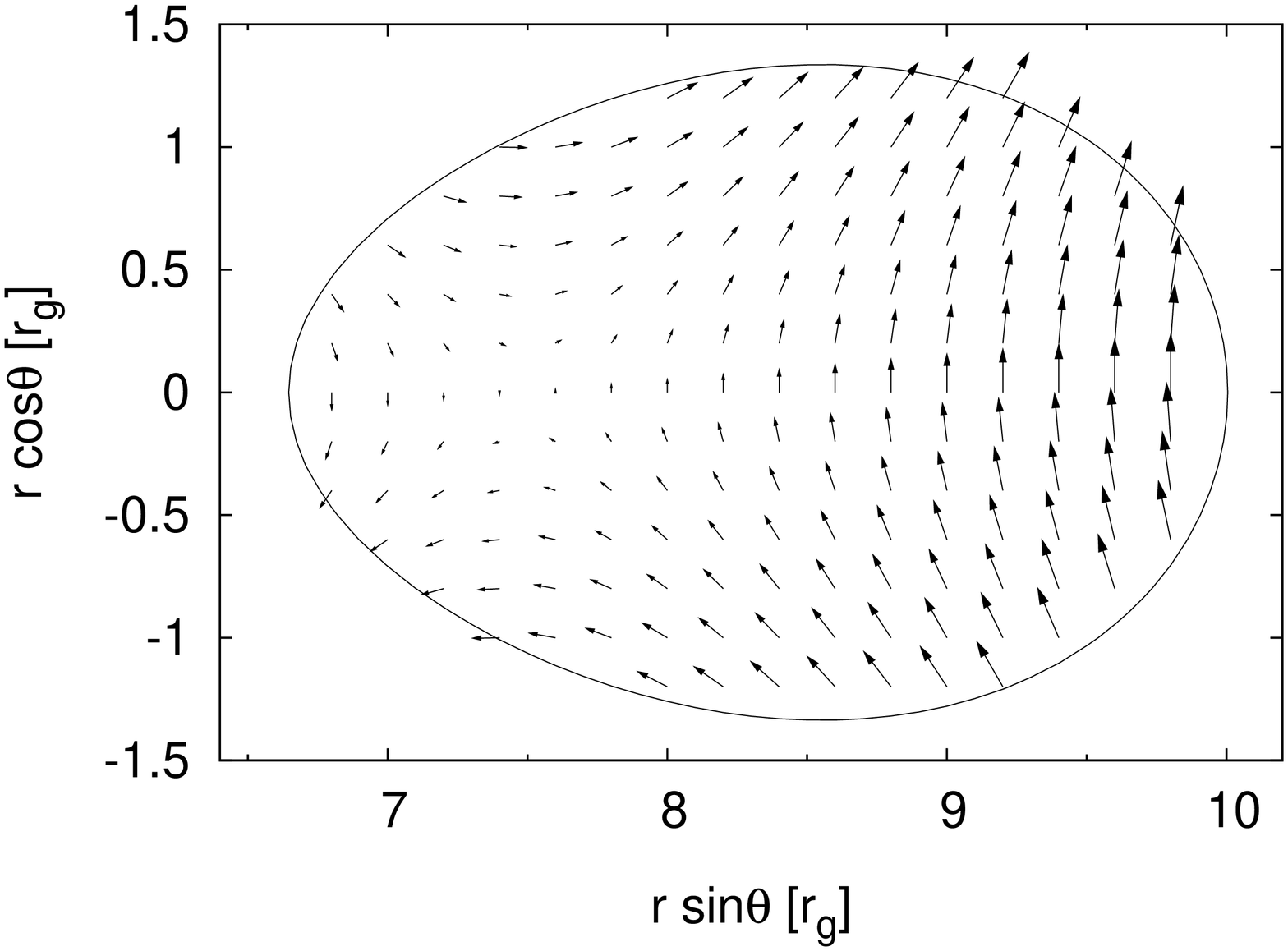}
\bigskip

\includegraphics[height=.48\textwidth,angle=-90]{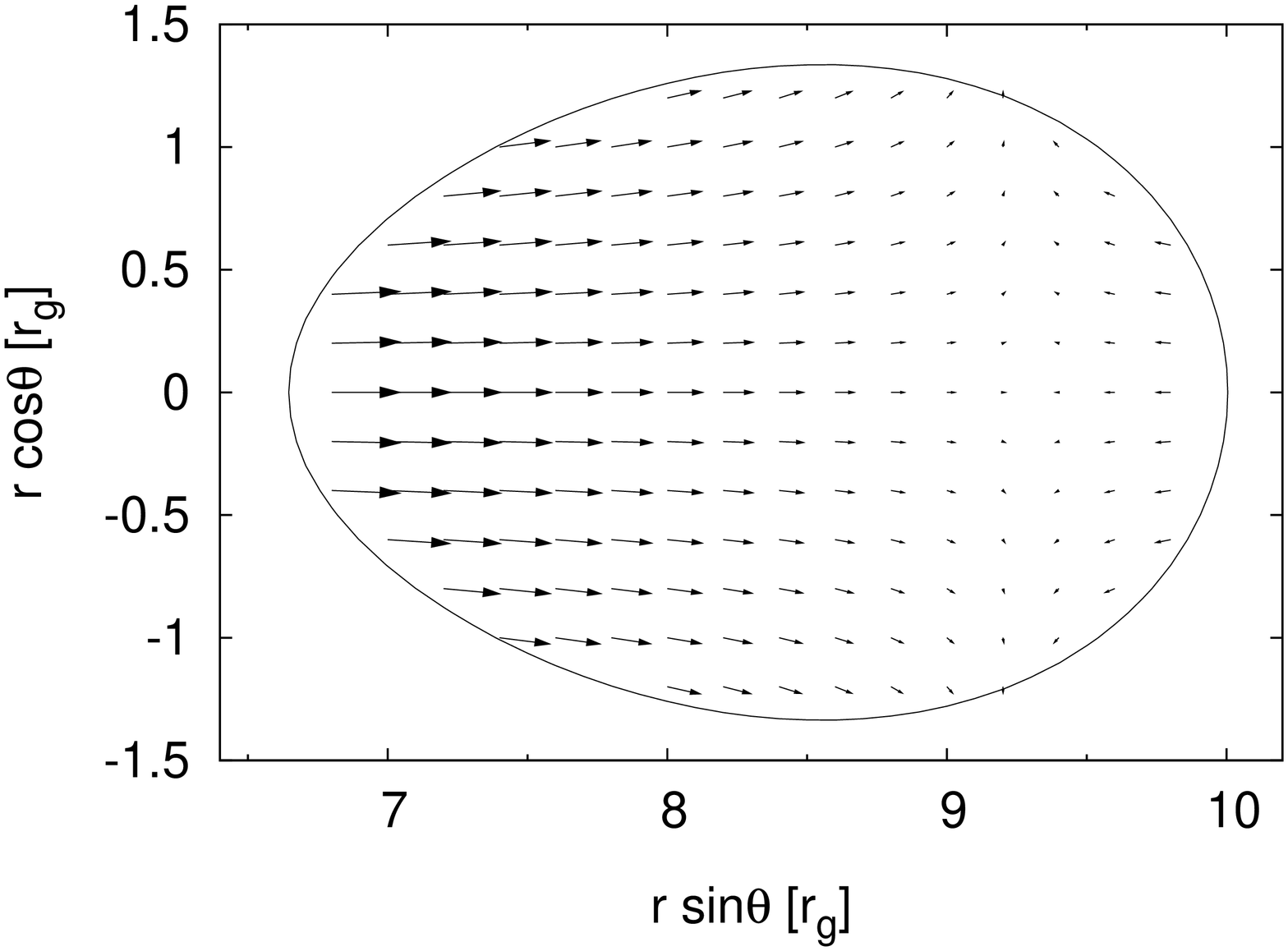} 
\hfill
\includegraphics[height=.48\textwidth,angle=-90]{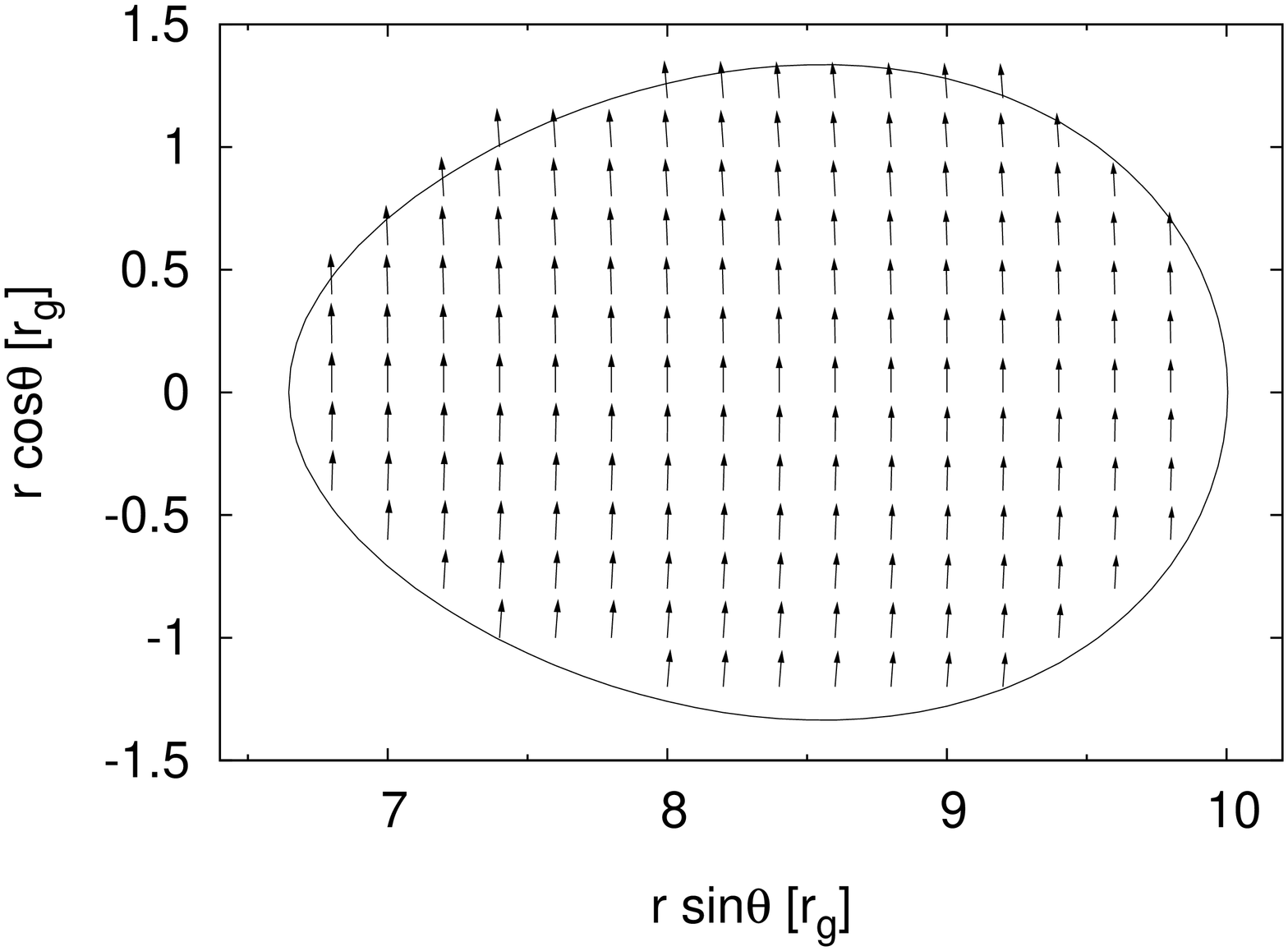}
\end{center}
\caption{\label{fig:velocity-nonaxi1} \label{fig:velocity-nonaxi2} Poloidal velocity fields of the non-axisymmetric epicyclic modes for the same torus as in \fref{fig:velocity-axi}. {\it Top\/} The $m=1$ radial ({\it left\/}) and vertical ({\it right\/}) epicyclic mode. {\it Bottom\/} 
The $m=-1$ radial ({\it left\/}) and vertical ({\it right\/}) epicyclic mode.}
\end{figure}
\end{center}

Poloidal velocity fields of the non-axisymmetric modes are displayed in figure \ref{fig:velocity-nonaxi1}.
The $m=1$ radial mode velocity field exhibits radial flows originating in the central regions of the torus and pointing outwards in opposite directions (top left panel). The $m=-1$ radial mode velocity field has a similar character, except that the radial flows point inwards (bottom left panel). The poloidal flow of the $m=1$ vertical mode has analogous features to that in the axisymmetric case (compare the top right panel of \fref{fig:velocity-nonaxi2} to the right panel of \fref{fig:velocity-axi}), while for the $m=-1$ vertical mode the flow is, apart from small variations, mainly vertical (\fref{fig:velocity-nonaxi2}, bottom right panel). Once again, the shape of the velocity fields described here remains preserved for all values of the black hole spin.

\bigskip

\citet{bla07} presented the properties of the axisymmetric and non-axisymmetric radial epicyclic mode frequencies for non-slender tori in a spherical pseudo-Newtonian potential, likewise based on calculations accurate to second-order with respect to the torus thickness. Comparing their figures  2, 6, 7 to our figures \ref{fig:radial}, \ref{fig:vertical}, \ref{fig:rapid-axi}, \ref{fig:radial-plus} and \ref{fig:rapid-nonaxi}, one may distinguish the behaviour of frequencies calculated in a pseudo-Newtonian potential from those calculated in a fully-relativistic Schwarzschild and Kerr geometry. Our results show that the radial and vertical mode frequencies in a rotating, axially symmetric (Kerr) spacetime follow the same trend as those calculated in a non-rotating, spherically symmetric potential, except that the vertical frequency has in a rotating spacetime a non-monotonic profile already in the case of test particles. This non-monotonicity feature is for thicker tori present in both, axially and spherically symmetric, potentials. Generally, only extremely high black hole spin values ($a\gtrsim 0.96$) cause slight variations in the frequency behaviour.

In order to compare the poloidal velocity fields of the axisymmetric modes in Kerr geometry (our \fref{fig:velocity-axi}) to those derived in the pseudo-Newtonian potential see figure 5 of \citet{bla07}.

\section{Conclusions}
\label{sec:conclusions}

We have assumed a pressure-supported torus of small radial extent in a Kerr spacetime that performs epicyclic oscillations and studied the pressure effects on the epicyclic modes properties, i.e., how the modes eigenfunctions and eigenfrequencies of a torus differ from those of a free test particle. For this purpose we calculated the relevant pressure corrections to the axisymmetric and lowest-order non-axisymmetric epicyclic mode eigenfunctions and eigenfrequencies within  first- (eigenfunctions) and second-order (eigenfrequencies) accuracy in torus thickness.

In the limit of an infinitely slender torus, the radial and vertical epicyclic oscillations occur as global oscillations that correspond to purely radial and vertical displacements of the whole torus at epicyclic frequencies of free test particles orbiting at the position of the torus pressure maximum \citep{abr06,bla06}. Several numerical studies \citep[e.g.][]{mon04,rub05a,rub05b,sra07} have shown that when the torus gets thicker, its (axisymmetric) oscillations occur at lower frequencies. This has been confirmed recently by analytic pseudo-Newtonian calculations \citep{bla07}, where pressure corrections to epicyclic modes of a small-size, constant specific angular momentum torus in the \citet{pac80} potential were derived. Using the same approach, but within the framework of general relativity, we extended their results into the rotating Kerr geometry. For both axisymmetric and the radial non-axisymmetric oscillations explored in \citet{bla07}, our calculations qualitatively confirm the trends as carried out there for a spherically symmetric potential. As expected (and also demonstrated in \citet{bla07}), the epicyclic mode eigenfunctions of thicker tori no longer describe a purely radial or vertical displacement since there appear some deviations of the flow in the poloidal velocity fields for both modes.

The configuration considered here is represented by the idealised model of a non-selfgravitating, non-accreting, non-magnetised torus with constant specific angular momentum\footnote{Uniform specific angular momentum distribution throughout the torus is (aside from simplicity reasons) assumed because the relativistic Papaloizou-Pringle equation for non-constant distributions does not describe a self-adjoint eigenvalue problem (see equation (26) in \citet{bla06}), and we do not have the appropriate complete orthonormal set of eigenfunctions that is necessary to apply the perturbation method (see also discussion in \citet{bla07}).}, studied within a purely hydrodynamical regime. These tori are widely known to be dynamically unstable under global, non-axisymmetric perturbations \citep{pap84}. However, this instability can be suppressed when accretion through the inner edge of the torus takes place \citep{bla87,bla88}. A possible interpretation is that torus-like structures can be formed within the innermost regions of a standard, nearly-Keplerian accretion disc where several physical processes may give rise to pressure gradients that in turn form tori. Such torus-like accretion flows seem to appear in three-dimensional global MHD simulations \citep{dev03,mac06}. 

Whether purely hydrodynamic modes may exist in the presence of magnetic fields that surely play a significant role in the physics of accretion flows is an issue still to be investigated more deeply. A few studies in that context have been carried out, e.g. by \citet{mon07} who explored the oscillation properties of relativistic tori comprising a toroidal magnetic field, and reported similar results as obtained in previous investigations of  non-magnetised torus oscillations. Their initial set-up, however, did not allow for development of the magneto-rotational turbulence (MRI). One of the first attempts to investigate the effects of MRI on the properties of hydrodynamic oscillation modes via relativistic MHD simulations was done by \citet{fra05}.

In general, studying oscillations of black hole accretion discs can improve our understanding of the origin of the observed X-ray variability. In particular HF QPOs that are detected in the X-ray light curves of several X-ray binaries are often attributed to disc oscillations. The results discussed here can be applied to models that directly involve epicyclic oscillations. Assuming a particular oscillation model, the identification of oscillation frequencies with the frequencies of observed QPOs can provide a precise determination of the mass or spin of the black hole \citep[e.g.][]{wag01,abr01,kat06}. Black hole spin estimates for several microquasars have been carried out, based on the resonance model considering epicyclic oscillations in a thin disc that occur at frequencies of free test particles \citep{tor05a}. Applying pressure corrections to epicyclic frequencies, our results should be taken into account to obtain (more) accurate estimations \citep{bla07}.

\ack

This work was supported by the Polish Ministry of Science grant N203 009 31/1466 (OS) and the Czech grant LC06014 (ES). The authors would like to thank Marek Abramowicz for initiating this work and his invaluable support, as well as W\l odek Klu\'{z}niak and Omer Blaes for their kind advice and encouragement. We also thank Pavel Bakala and Gabriel T{\"o}r{\"o}k for many useful discussions and technical help. Then we would like to acknowledge the hospitality of the Silesian University in Opava and NORDITA in Copenhagen and Stockholm where most of the work was carried out.

\section*{Appendix}
\label{sec:appendix}
\appendix
\setcounter{section}{1}


Here we display the coefficients introduced to abbreviate the analytic terms we derive and use in section \ref{sec:calculation}.

\subsection{The first-order terms}

The coefficients that appear in the first-order expansion terms of $A$, $\bar{\Omega}$, $f$, $\hat{L}$ in subsection \ref{sec:first} read
\bea
\mathcal{A}_{11} &=& -2L_{101}\frac{r_0^{1/2}(a^2-2ar_0^{1/2}+r_0^2)}{(r_0-a^2)(r_0^{3/2}+a)},\\
\Omega_{11} &=& -L_{101}\frac{r_0^2(2a-3r_0^{1/2}+r_0^{3/2})}{(r_0-a^2)(r_0^{3/2}+a)},
\eea
\bea
f_{11}&=&\frac{2}{r_0^{4}} \left[\frac{r_0^2}{a^2 + (r_0-2) r_0}\right]^{1/2}\left\{-a^4 + 2 a^3 r_0^{1/2} - 4 a^2 (r_0-1) r_0\right.\nn \\
&& \left. + 2 a r_0^{3/2} (5 r_0-6) + r_0^2 [8 + (r_0-8) r_0]\right\},
\eea
\be
f_{12}= \frac{2}{r_0^{4}} \left[\frac{r_0^2}{a^2 - 2 r_0 + r_0^2}\right]^{1/2} \left(3 a^2 - 4 a r_0^{1/2} + r_0^2\right)\left(a^2 - 2 r_0 + r_0^2\right),
\ee
\bea
L_{101} &=& \frac{2(r_0 - a^2)}{r_0^2} \left[\frac{r_0^2}{a^2+r_0(r_0-2)}\right]^{1/2},\\
L_{102} &=& -L_{101} \R^2+f_{11},\\
L_{103} &=& -L_{101} \T^2+f_{12},\\
L_{104} &=& -2\left[\frac{r_0^2}{a^2+r_0(r_0-2)}\right]^{-1/2},\\
L_{105} &=& -L_{104}\R^2+f_{11},\\
L_{106} &=& -L_{104}\T^2+f_{12},\\
L_{107} &=& \frac{r_0(r_0 - 1)}{(r_0-a^2)} L_{101},\\
L_{108} &=& -(2nL_{101}+L_{107})\R^2 + 3n f_{11},\\ 
L_{109} &=& -L_{107}\T^2 + nf_{12},\\
L_{110} &=& 2n(f_{12}-L_{104}\T^2).
\eea

\subsection{The second-order terms}

The coefficients in the second-order terms of $A$, $\bar{\Omega}$, $f$, $\hat{L}$ in subsection \ref{sec:second} take the following forms
\bea
\mathcal{A}_{21} &=& \frac{1}{\left[2a+\left(r_0-3\right)r_0^{1/2}\right]\left[a^2+ \left(r_0-2\right)r_0\right]\left(a+r_0^{3/2}\right)^2\,r_0^{3/2}}\times\nn \\
\bigskip
&&\times\left[5a^6+2a^5\left(3r_0-2\right)r_0^{1/2}+a^4\left[\left(r_0+10\right)r_0-64\right]r_0\right.\nn \\
&&\left.+4a^3\left[\left(6r_0-23\right)r_0+40\right]r_0^{3/2}+a^2\left\{\left[\left(6r_0-67\right)r_0+100\right]r_0\right.\right.\nn \\
&&\left.\left.-108\right\}r_0^2+2a\left(r_0+16\right)r_0^{9/2}+\left(5r_0-16\right)r_0^6\right],
\eea
\bea
\mathcal{A}_{22} &=& -\frac{5a^3+a^2(r_0-8)r_0^{1/2}+3ar_0^2-r_0^{7/2}}{[2a+ (r_0-3)r_0^{1/2}](a+r_0^{3/2})r_0^{3/2}},
\eea
\bea
\Omega_{21} &=& \frac{a+(r_0-2)r_0^{1/2}}{[a^2+(r_0-2)r_0](a+r_0^{3/2})^2}\left[ a^3-a^2(r_0+6)r_0^{1/2}\right.\nn\\
&&\left.+3a(3r_0+2)r_0+3(r_0-4)r_0^{5/2}\right],
\eea
\be
\Omega_{22} = 1-\frac{2a}{a+r_0^{3/2}},
\ee
\bea
f_{21}&=&\frac{1}{2 \left[2 a + (r_0-3) r_0^{1/2}\right] r_0^4 \left[a^2 + (r_0-2) r_0\right]}\times\left[8 a^7 \right.\nn\\
&& \left. + 4 a^5 r_0 (3 r_0+8) + a^6 r_0^{1/2} (4 r_0-37) + a^4 r_0^{3/2} \left[5 r_0 (2 r_0-17)+74\right]\right.\nn\\
&& \left. + 16 a^3 r_0^2 \left[r_0 (3 r_0-1)-5\right] + 4 a r_0^3 \left\{r_0 [(124 - 21 r_0) r_0-192]+88\right\}\right.\nn\\
&& \left. + a^2 r_0^{5/2} \left\{r_0 \left[r_0 (32 r_0-307)+500\right]-188\right\} \right.\nn\\
&& \left. + r_0^{7/2} \left(r_0 \left\{r_0 \left[(77 - 6 r_0) r_0-286\right]+380\right\}-168\right)\right],
\eea
\bea
f_{22}&=&-\frac{1}{r_0^4 (2 a - 3 r_0^{1/2} + r_0^{3/2})}\times\left(24 a^5 - 85 a^4 r_0^{1/2} + 72 a^3 r_0 \right.\nn\\
&& \left. + 34 a^2 r_0^{3/2} + 12 a^4 r_0^{3/2} - 48 a r_0^2 + 2 a^3 r_0^2 - 69 a^2 r_0^{5/2} + 52 a r_0^3 \right.\nn\\ 
&& \left. + 12 r_0^{7/2} + 11 a^2 r_0^{7/2} - 6 a r_0^4 - 14 r_0^{9/2} + 3 r_0^{11/2}\right),
\eea
\bea
f_{23}&=&\frac{1}{6 r_0^4 \left(2 a - 3 r_0^{1/2} + r_0^{3/2}\right)}\times\left(24 a^5 - 63 a^4 r_0^{1/2} + 24 a^3 r_0 + 24 a^2 r_0^{3/2}\right.\nn\\
&& \left. + 12 a^4 r_0^{3/2} - 10 a^2 r_0^{5/2} - 24 a r_0^3 + 8 a r_0^4 +  9 r_0^{9/2} - 4 r_0^{11/2}\right),
\eea
\bea
L_{201} &=& f_{21} + L_{101}f_{11} - L_{202}\R^2,\\
L_{202} &=& \frac{3a^2-2r_0}{r_0^2},\\
L_{203} &=& f_{22}+L_{101}f_{12}-L_{202}\T^2-L_{204}\R^2,\\
L_{204} &=& -\frac{a^2}{r_0^2},\\
L_{205} &=& f_{23}-L_{204}\T^2,\\
\bigskip
L_{206} &=& f_{21}+(L_{101}-L_{107})f_{11}-L_{207}\R^2,\\
L_{207} &=& \frac{3[a^2+(r_0-2)r_0]}{r_0^2},\\
L_{208} &=& f_{22}+(L_{101}-L_{107})f_{12}-L_{207}\T^2-L_{204}\R^2,\\
L_{209} &=& f_{23}-L_{204}\T^2,\\
\bigskip
L_{210} &=& L_{101}\left(3n+\frac{r_0^2-r_0}{r_0-a^2}\right)f_{11}-\left[\frac{6na^2- 2r_0(r_0+2n-2)}{r_0^2}\right]\R^2\nn \\
&&+4nf_{21},\\
L_{211} &=& -2+\frac{4}{r_0},\\
L_{212} &=& 2nf_{22}+L_{101}\left(n+\frac{r_0^2-r_0}{r_0-a^2}\right)f_{12} - L_{211}\T^2 - 2nL_{204}\R^2,\\
\bigskip
L_{213} &=& 2n\left[f_{22}+(L_{101}-L_{107})f_{12} -L_{207}\T^2\right]+\R^2,\\
L_{214} &=& -1,\\
L_{215} &=& 4nf_{23}-(2nL_{204}-1)\T^2,\\
\bigskip
L_{216} &=& \frac{r_0}{a^2 + r_0(r_0-2)}\left\{[r_0^3 +a^2(r_0+2)] \left(\omega_i^{(0)}\Omega_0\right)^2 \right.\\
&& \left. -{m^2(r_0-2)-4ma\omega_i\x{0}\Omega_0}\right\},\\
L_{217} &=& - L_{216}\R^2,\\
L_{218} &=& - L_{216}\T^2.
\eea

\subsection{Eigenfunctions-related coefficients}

The coefficients $w_{41}$, $w_{42}$, $w_{51}$, $w_{52}$ introduced in table \ref{tab:eigenfunctions} have the form

\bea
w_{41}=-\frac{\R^2 (2 \T^2 + 2 n \T^2 - n \sigma_4^2)}{\T^2 - \R^2},\\
w_{42}=\frac{\T^2 (2 \R^2 + 2 n \R^2 - n \sigma_4^2)}{\T^2 - \R^2},\\
w_{51}=-\frac{\R^2 (2 \T^2 + 2 n \T^2 - n \sigma_5^2)}{\T^2 - \R^2},\\
w_{52}=\frac{\T^2 (2 \R^2 + 2 n \R^2 - n \sigma_5^2)}{\T^2 - \R^2}.
\eea

\References

\bibitem[{{Abramowicz and Klu{\'z}niak}(2001)}]
{abr01} Abramowicz M A and Klu{\'z}niak W 2001 {\it Astron. Astrophys.\/} {\bf 374\/} L19

\bibitem[{{Abramowicz \etal}(2006)}]
{abr06} Abramowicz M A, Blaes O M, Hor\'ak J, Klu{\'z}niak W and Rebusco P 2006 \CQG {\bf 23\/} 1689

\bibitem[{{Aliev and Galtsov}(1981)}]
{ali81} Aliev A N and Galtsov D V 1981 {\it Gen. Rel. Grav. \/} {\bf 13\/} 899

\bibitem[{{Blaes}(1985)}]
{bla85} Blaes O M 1985 {\it Mon. Not. R. Astron. Soc.\/} {\bf 216\/} 553

\bibitem[{{Blaes}(1987)}]
{bla87} Blaes O M 1987 {\it Mon. Not. R. Astron. Soc.\/} {\bf 227\/} 975

\bibitem[{{Blaes and Hawley}(1988)}]
{bla88} Blaes O M and Hawley J F 1988 {\it Astrophys. J. \/} {\bf 326\/} 277

\bibitem[{{Blaes, Arras and Fragile}(2006)}]
{bla06} Blaes O M, Arras P and Fragile P C 2006 {\it Mon. Not. R. Astron. Soc.\/} {\bf 369\/} 1235

\bibitem[{{Blaes \etal}(2007)}]
{bla07} Blaes O M, {\v S}r{\'a}mkov{\'a} E, Abramowicz M A, Klu{\'z}niak W and Torkelsson U 2007 {\it ApJ\/} {\bf 665\/} 642

\bibitem[{{De Villiers and Hawley \etal}(2003)}]
{dev03} De Villiers J P and Hawley J F 2003  {\it Astrophys. J. \/} {\bf 665\/} 642

\bibitem[{{Fragile}(2005)}]
{fra05} Fragile P C 2005 {\it eprint\/} arXiv:astro-ph/0503305

\bibitem[{{Fragile \etal}(2008)}]
{fra08} Fragile P C, Lindner C C, Anninos P, Salmonson J D 2008 {\it eprint \/} arXiv:0809.3819

\bibitem[{{Kato \etal}(1998)}]
{kat98} Kato S, Fukue J and Mineshige S 1998 {\it Kyoto Univ. press\/} Kyoto 1998

\bibitem[{{Kato}(2001a)}]
{kat01a} Kato S 2001a {\it Publ. Astron. Soc. Japan\/} {\bf 53\/} 1

\bibitem[{{Kato}(2001b)}]
{kat01b} Kato S 2001b {\it Publ. Astron. Soc. Japan\/} {\bf 53\/} 37

\bibitem[{{Kato}(2003)}]
{kat03} Kato S 2003 {\it Publ. Astron. Soc. Japan\/} {\bf 55\/} 801

\bibitem[{{Kato}(2004)}]
{kat04} Kato S 2004 {\it Publ. Astron. Soc. Japan\/} {\bf 56\/} 905

\bibitem[{{Kato and Fukue}(2006)}]
{kat06} Kato S and Fukue J 2006 {\it Publ. Astron. Soc. Japan\/} {\bf 58\/} 909

\bibitem[{{van der Klis}(2004)}]
{van04} van der Klis M 2004  {\it eprint\/} arXiv:astro-ph/0410551

\bibitem[{{Klu{\'z}niak and Abramowicz}(2000)}]
{klu00} Klu{\'z}niak W and Abramowicz M A 2000 submitted to {\it Phys. Rev. Lett. \/} {\it eprint\/} arXiv:astro-ph/0105057

\bibitem[{{Kojima}(1986)}]
{koj86} Kojima Y 1986 {\it Prog. Theor. Phys.\/} {\bf 75\/} L1464

\bibitem[{{Machida \etal}(2006)}]
{mac06} Machida M, Nakamura K E and Matsumoto R 2006 {\it Publ. Astron. Soc. Japan\/} {\bf 58\/} 193

\bibitem[{{McClintock and Remillard}(2003)}]
{mcc03} McClintock J E and Remillard R A 2003  {\it eprint\/} arXiv:astro-ph/0306213

\bibitem[{{Montero \etal}(2007)}]
{mon07} Montero P J, Zanotti O, Font J A and Rezzolla L 2007  {\it Mon. Not. R. Astron. Soc.\/} {\bf 378\/} 1101

\bibitem[{{Montero \etal}(2004)}]
{mon04} Montero P J, Rezzolla L, Yoshida S 2004 {\it Mon. Not. R. Astron. Soc.\/} {\bf 354\/} 1040

\bibitem[{{Nowak and Lehr}(1998)}]
{now98} Nowak M A and Lehr D E 1998 {\it Theory of black hole accretion disks\/} ed M A Abramowicz, G Bj{\"o}rnsson and J E Pringle (Cambridge University Press) p 233

\bibitem[{{Paczy{\'n}ski and Wiita}(1980)}]
{pac80} Paczy{\'n}ski B and Wiita P J 1980 {\it Astron. Astrophys.\/} {\bf 88\/} 23

\bibitem[{{Papaloizou \& Pringle}(1984)}]
{pap84} Papaloizou J C B and Pringle J E 1984 {\it Mon. Not. R. Astron. Soc.\/} {\bf 208\/} 721

\bibitem[{{Proga \& Begelman}(2003)}]
{pro03a} Proga D and Begelman 2003 {\it Astrophys. J.\/} {\bf 583\/} 69

\bibitem[{{Rezzolla \etal}(2003a)}]
{rez03a} Rezzolla L, Yoshida S and Zanotti O  2003a {\it Mon. Not. R. Astron. Soc.\/} {\bf 344\/} 978

\bibitem[{{Rezzolla \etal}(2003b)}]
{rez03b} Rezzolla L, Yoshida S, Maccarone T J and Zanotti O  2003b {\it Mon. Not. R. Astron. Soc.\/} {\bf 344\/} 37

\bibitem[{{Rubio-Herrera and Lee}(2005a)}]
{rub05a} Rubio-Herrera E and Lee W H 2005a  {\it Mon. Not. R. Astron. Soc.\/} {\bf 357\/} L31

\bibitem[{{Rubio-Herrera and Lee}(2005b)}]
{rub05b} Rubio-Herrera E and Lee W H 2005b  {\it Mon. Not. R. Astron. Soc.\/} {\bf 362\/} 789

\bibitem[{{{\v S}r{\'a}mkov{\'a} \etal}(2007)}]
{sra07} {\v S}r{\'a}mkov{\'a} E, Torkelsson U and Abramowicz M A  2007  {\it Astron. Astrophys.\/} {\bf 467\/} 641

\bibitem[{{Stella and Vietri}(1998)}]
{ste98} Stella L and Vietri M 1998 {\it Astrophys. J. Lett. \/} {\bf 492\/} L59 

\bibitem[{{T{\"o}r{\"o}k \etal}(2005)}]
{tor05a} T{\"o}r{\"o}k  G, Abramowicz M A, Klu{\'z}niak W and Stuchl{\'i}k Z 2005 {\it Astron. Astrophys. \/} {\bf 436\/} 1

\bibitem[{{T{\"o}r{\"o}k and Stuchl{\'i}k}(2005)}]
{tor05b} T{\"o}r{\"o}k  G and Stuchl{\'i}k Z 2005 {\it Astron. Astrophys. \/} {\bf 437\/} 775

\bibitem[{{Wagoner}(1999)}]
{wag99} Wagoner R V 1999 {\it Phys. Rev.\/} {\bf 311} 259

\bibitem[{{Wagoner \etal}(2001)}]
{wag01} Wagoner R V, Silbergleit A S and Ortega-Rodr{\'i}guez M 2001 {\it Astrophys. J. Lett. \/} {\bf 559\/} L25

\bibitem[{{Zanotti \etal}(2003)}]
{zan03} Zanotti O, Rezzolla L and Font J A 2003 {\it Mon. Not. R. Astron. Soc.\/} {\bf 341\/} 832

\endrefs

\end{document}